\documentclass[twocolumn]{aastex631}
\usepackage{lipsum}
\usepackage{breqn}
\usepackage{amsmath}
\usepackage{gensymb}
\usepackage{hyperref}

\usepackage{ulem}

\hypersetup{
    colorlinks=true,
    linkcolor=blue,
    }

\newcommand{\HI}{\ion{H}{1}~}
\newcommand{\HII}{\ion{H}{2}~}
\newcommand{\kms}{km~s$^{-1}~$}
\newcommand{\Lya}{Ly$\alpha$ }

\newcommand{\Ha}{H$\alpha$ }
\newcommand{\ngc}{NGC 99 }

\received{March 28, 2024}
\revised{July 24, 2024}
\accepted{August 11, 2024}

\submitjournal{ApJ}

\shorttitle{Indirect Evidence of Inflows in NGC 99}
\shortauthors{Olvera et al.}
\graphicspath{{./}{figures/}}

\begin{document}

\title{DIISC-IV: DIISCovery of Anomalously Low Metallicity \HII Regions in NGC 99: \\
Indirect Evidence of Gas Inflows}

\correspondingauthor{Alejandro J. Olvera}
\email{ajolver2@asu.edu}

\author[0000-0002-2819-0753]{Alejandro J. Olvera}
\affiliation{School of Earth and Space Exploration, Arizona State University, 781 Terrace Mall, Tempe, AZ 85287, USA}

\author[0000-0002-2724-8298]{Sanchayeeta Borthakur}
\affiliation{School of Earth and Space Exploration, Arizona State University, 781 Terrace Mall, Tempe, AZ 85287, USA}

\author[0000-0002-3472-0490]{Mansi Padave}
\affiliation{School of Earth and Space Exploration, Arizona State University, 781 Terrace Mall, Tempe, AZ 85287, USA}

\author[0000-0001-6670-6370]{Timothy Heckman}
\affiliation{School of Earth and Space Exploration, Arizona State University, 781 Terrace Mall, Tempe, AZ 85287, USA}
\affiliation{Department of Physics \& Astronomy, Johns Hopkins University, Baltimore, MD 21218, USA}

\author[0000-0003-1436-7658]{Hansung B. Gim}
\affiliation{School of Earth and Space Exploration, Arizona State University, 781 Terrace Mall, Tempe, AZ 85287, USA}
\affiliation{Department of Physics, Montana State University, P. O. Box 173840, Bozeman, MT 59717, USA}

\author[0000-0001-5530-2872]{Brad Koplitz}
\affiliation{School of Earth and Space Exploration, Arizona State University, 781 Terrace Mall, Tempe, AZ 85287, USA}

\author[0000-0003-1739-3640]{Christopher Dupuis}
\affiliation{School of Earth and Space Exploration, Arizona State University, 781 Terrace Mall, Tempe, AZ 85287, USA}

\author[0000-0003-3168-5922]{Emmanuel Momjian}
\affiliation{National Radio Astronomy Observatory, 1003 Lopezville Road, Socorro, NM 87801, USA}

\author[0000-0003-1268-5230]{Rolf A. Jansen}
\affiliation{School of Earth and Space Exploration, Arizona State University, 781 Terrace Mall, Tempe, AZ 85287, USA}



\begin{abstract}
As a part of the Deciphering the Interplay between the Interstellar medium, Stars, and the Circumgalactic medium (DIISC) survey, we investigate indirect evidence of gas inflow into the disk of the galaxy \object{NGC 99}. We combine optical spectra from the Binospec spectrograph on the MMT telescope with optical imaging data from the Vatican Advanced Technology Telescope, radio \HI 21\,cm emission images from the NSF Karl G. Jansky's Very Large Array, and UV spectroscopy from the Cosmic Origins Spectrograph on the Hubble Space Telescope. We measure emission lines (H$\alpha$, H$\beta$, [\ion{O}{3}]$\lambda5007$, [\ion{N}{2}]$\lambda6583$, and [\ion{S}{2}]$\lambda6717,31$) in 26 \HII regions scattered about the galaxy and estimate a radial metallicity gradient of $-0.017$ dex kpc$^{-1}$ using the N2 metallicity indicator. Two regions in the sample exhibit an anomalously low metallicity (ALM) of 12+log(O/H) = 8.36 dex, which is $\sim$0.16 dex lower than other regions at that galactocentric radius. They also show a high difference between their \HI and \Ha line of sight velocities on the order of 35 km s$^{-1}$. Chemical evolution modeling indicates gas accretion as the cause of the ALM regions. We find evidence for corotation between the interstellar medium of \object{NGC 99} and Ly$\alpha$ clouds in its circumgalactic medium, which suggests a possible pathway for low metallicity gas accretion. We also calculate the resolved Fundamental Metallicity Relation (rFMR) on sub-kpc scales using localized gas-phase metallicity, stellar mass surface density, and star-formation rate surface density. The rFMR shows a similar trend as that found by previous localized and global FMR relations. 
\end{abstract}



\section{Introduction} \label{sec:intro}
Cosmological simulations confirm that gas cycling in and out of galaxies, also known as the baryon cycle, is crucial to their growth. Gas inflows into galactic disks are necessary to maintain star formation over cosmic time, whereas galactic outflows are important to regulating star formation rates \citep{Keres_2005MNRAS.363....2K, Hopkins_2014MNRAS.445..581H, Faucher_2023ARA&A..61..131F}. The interstellar medium (ISM) serves as the center stage for the baryon cycle as it is where accreted gas may collapse to form stars and where outflows from stellar remnants are produced. Thus, in order to further understand the physical mechanisms affecting the ISM, it is vital to know its atomic gas content ($\rm M_{HI}$), molecular gas content ($\rm M_{H_{2}}$), star-formation rate (SFR), and metallicity (Z). 

Measurement of metal abundances in galaxies provides an observational approach to inferring gas in/out flows and star formation since most metals are the result of the nucleosynthesis within stars and in their supernovae explosions \citep{Fin_Dave_2008MNRAS.385.2181F,Lilly_2013ApJ...772..119L,Belfiore_2016MNRAS.455.1218B}. Therefore, the radial metallicity gradient of galaxies is of great interest to study how galaxies grow (e.g., inside-out or outside-in; \citealt{Larson_1976MNRAS.176...31L,Matteucci_1989MNRAS.239..885M, Pezzulli_2016MNRAS.455.2308P,Acharyya_2020MNRAS.495.3819A, Wang_2022ApJ...929...95W}). In the local universe, most massive galaxies show a negative metallicity gradient with metallicity dropping with galactocentric radius \citep{Vila-Costas_1992MNRAS.259..121V,Oey_1993ApJ...411..137O,Sanchez_2020ARA&A..58...99S}. Studies with large samples of galaxies such as the Calar Alto Legacy Integral Field Area (CALIFA) survey or the Sloan Digital Sky Survey (SDSS) hav
e measured a characteristic gradient of approximately -0.1 dex/$R_e$ from their distribution of slopes, where $R_e$ is the disk effective radius \citep{Sanchez_2014_slope_A&A...563A..49S,Sanchez-Men_2016A&A...587A..70S,Parikh_2021MNRAS.502.5508P}. However, individual galaxies can have multiple metallicity slopes within the disk of the galaxy \citep{San_men_multi_grad_2018A&A...609A.119S}.

In our current understanding of galaxy evolution, gas from the regions surrounding galaxies, namely, the circumgalactic medium (CGM) or intergalactic medium, is accreted onto the outer parts of the galactic disk \citep{Roskar_2010MNRAS.408..783R,Font_2011MNRAS.416.2802F,Moran_2012ApJ...745...66M,Lackner_2012MNRAS.425..641L, Sanchez_2014A&ARv..22...71S}. The accretion process is thought to be crucial to supplying galaxies with gas to continue forming stars. However, directly observing active gas accretion is challenging due to the low densities in the CGM.

Indirect evidence of gas accretion can be found via observations of the bright star-forming regions within galactic disks since inflows of ex-situ gas can have a chemical composition different from that of disk gas. If the newly introduced gas is metal-poor, it can dilute the metal content of the star-forming regions \citep{Tinsley_1973ApJ...186...35T, Bournaud_2009ApJ...694L.158B, Dekel_2009Natur.457..451D, Vincenzo_2016MNRAS.458.3466V, Pace_2021ApJ...908..165P}. The diluted regions can then become anomalies in the metallicity pattern across the galaxy. Several studies of individual galaxies have found these anomalously low metallicity (ALM) regions \citep{Sanchez_Alm_2014A&ARv..22...71S,Howk_a_2018ApJ...856..166H,Howk_b_2018ApJ...856..167H, Luo_2021ApJ...908..183L, Ju_2022ApJ...938...96J}. Recently, integral-field surveys with more statistical robustness, like the Mapping Nearby Galaxies at Apache Point Observatory (MaNGA) survey, have found a large sample of ALM regions wi
th a preference for low mass galaxy-hosts ($<10^{10} \rm M_\odot$) \citep{Hwang_2019ApJ...872..144H}. Theory and a growing amount of observations suggest that these ALM regions are a result of low Z gas accretion into the disk that form young stars \citep{Hwang_2019ApJ...872..144H, Howk_a_2018ApJ...856..166H, Scholz_2021MNRAS.505.4655S}.

\subsection{NGC 99}\label{subsec:ngc99}
As a pilot study, we selected \object{NGC 99} to explore radial metallicity gradients, search for ALM \HII regions, and investigate a resolved fundamental metallicity relation (FMR). \object{NGC 99} is a star-forming spiral galaxy at a distance of 79.4 Mpc. As part of the Deciphering the Interplay between the Interstellar medium, Stars, and the Circumgalactic medium (DIISC) survey (Borthakur et al. in prep), \object{NGC 99} has extensive multiwavelength data coverage allowing us to probe the chemical compositions of its star-forming regions, ISM, and CGM and connect that information to other physical properties such as stellar mass ($\rm M_{\star}$) and SFR. To do this, we combine data from the Very Large Array, the Vatican Advanced Technology Telescope, the Hubble Space Telescope, the Large Binocular Telescope, and our newly obtained multi-object spectra from the MMT. \cite{Padave_2024ApJ...960...24P} recently measured the global stellar mass ($\rm M_{\star,global}$) and sta
r-formation rate ($\rm SFR_{global}$) of \object{NGC 99} to be $4.17\times10^{10}~\rm M_\odot$ and 2.45~$\rm M_\odot~yr^{-1}$, respectively. Gim et al. (in prep) found the total \HI mass to be $1.68\times10^{10}~\rm M_\odot$. Other properties of \object{NGC 99} can be found in Table \ref{tab:galaxy properties}.

We combine optical spectra of 26 \HII regions with optical band and radio \HI 21\,cm imaging to study the disk of \object{NGC 99}. We also probe the CGM of the galaxy at 159 kpc from its center with a UV-bright QSO sightline. The paper is organized as follows. In Section \ref{sec:data}, we describe the suite of data available and its reduction. In Section \ref{sec:results}, we analyze radial gradients of the dust content and metallicity. In Section \ref{sec:discussion}, we discuss our results in the context of other studies, explore chemical evolution models, and derive a resolved FMR. We summarize our findings in Section \ref{sec:conclusion}.

\section{Observations and Data Analysis} \label{sec:data} 

\begin{figure}[]
\plotone{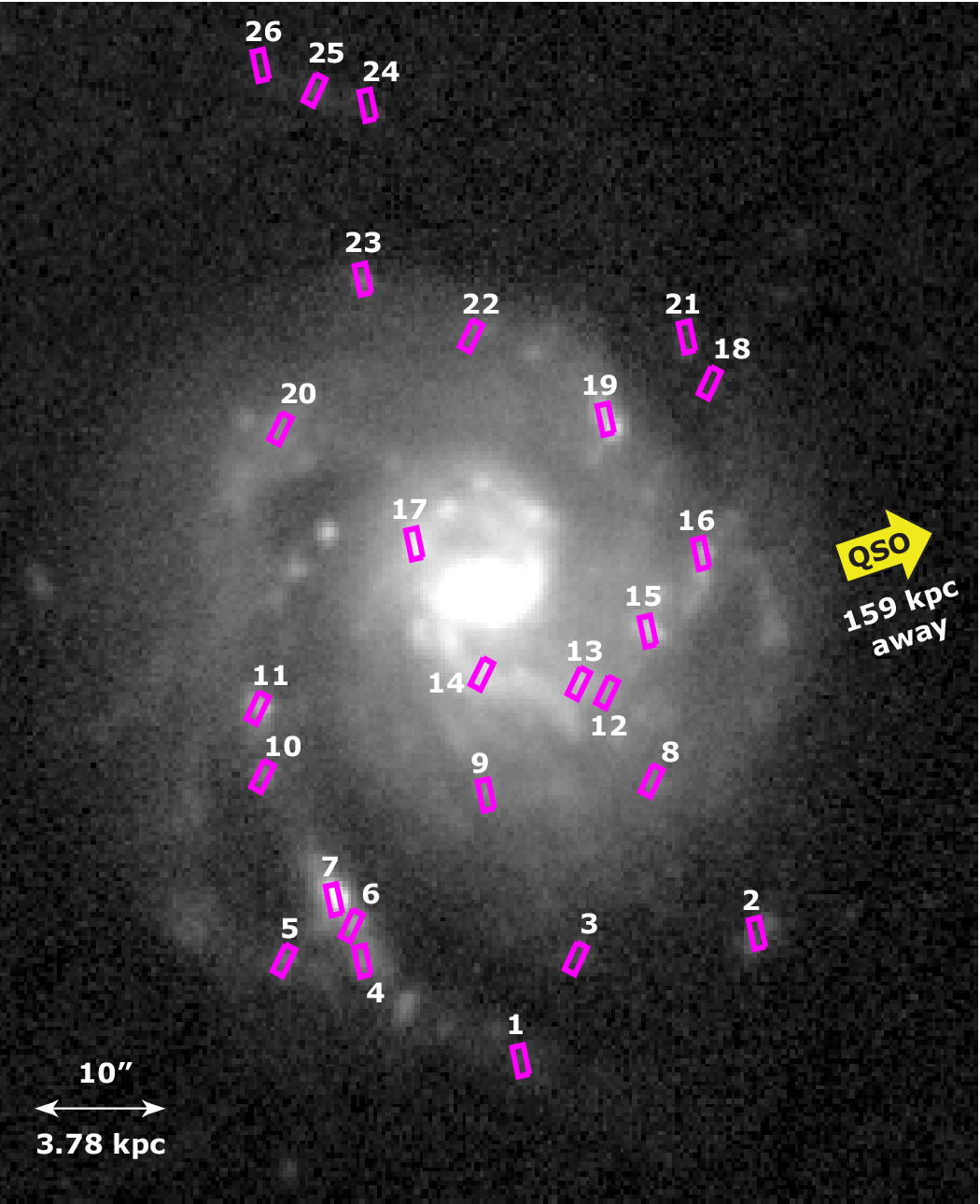}
\caption{A r-band image of NGC 99 obtained at the Vatican Advanced Technology Telescope with all the included Binospec slits placed on the \HII regions labeled by their assigned number. The yellow arrow points in the direction of QSO SDSS J002330.58+154744.9, which is a projected distance of 159 kpc away from the galaxy center.}
\label{fig:HII_regions}
\end{figure}

\subsection{Optical Spectroscopy of \HII Regions}\label{subsec:sample}
We obtained multi-slit spectra for \object{NGC 99} using the new spectroscopy instrument, Binospec, on the MMT 6.5m telescope located on Mt. Hopkins in Arizona. The observations were taken over a period of 4 nights between September 2020 and November 2021 with an average seeing of $1.15^{\prime\prime}$. A catalog of far-ultraviolet (FUV) bright targets was created using \texttt{SExtractor} \citep{Bertin_1996A&AS..117..393B} on an archival FUV image from the Galaxy Evolution Explorer (GALEX; \citealt{Martin_GALEX_2005ApJ...619L...1M,Morrissey_2007ApJS..173..682M}). The image is presented in \cite{Padave_2024ApJ...960...24P}. Regions with a FUV magnitude brighter than 25 were added to the catalog. UV regions are generally spatially correlated with H$\alpha$ regions as both are used as star-formation indicators. Hence, we are targeting and detecting \HII regions. The \texttt{BinoMask} software \citep{kansky_2019PASP..131g5005K} then automatically selected targets from the catalo
g by optimizing for the highest number of targets per mask design and placing slits to avoid effects of differential atmospheric refraction. \texttt{BinoMask} selected a total of 31 UV-bright \HII regions to observe.

\begin{figure*}[]
    \centering
    \includegraphics[scale=0.5]{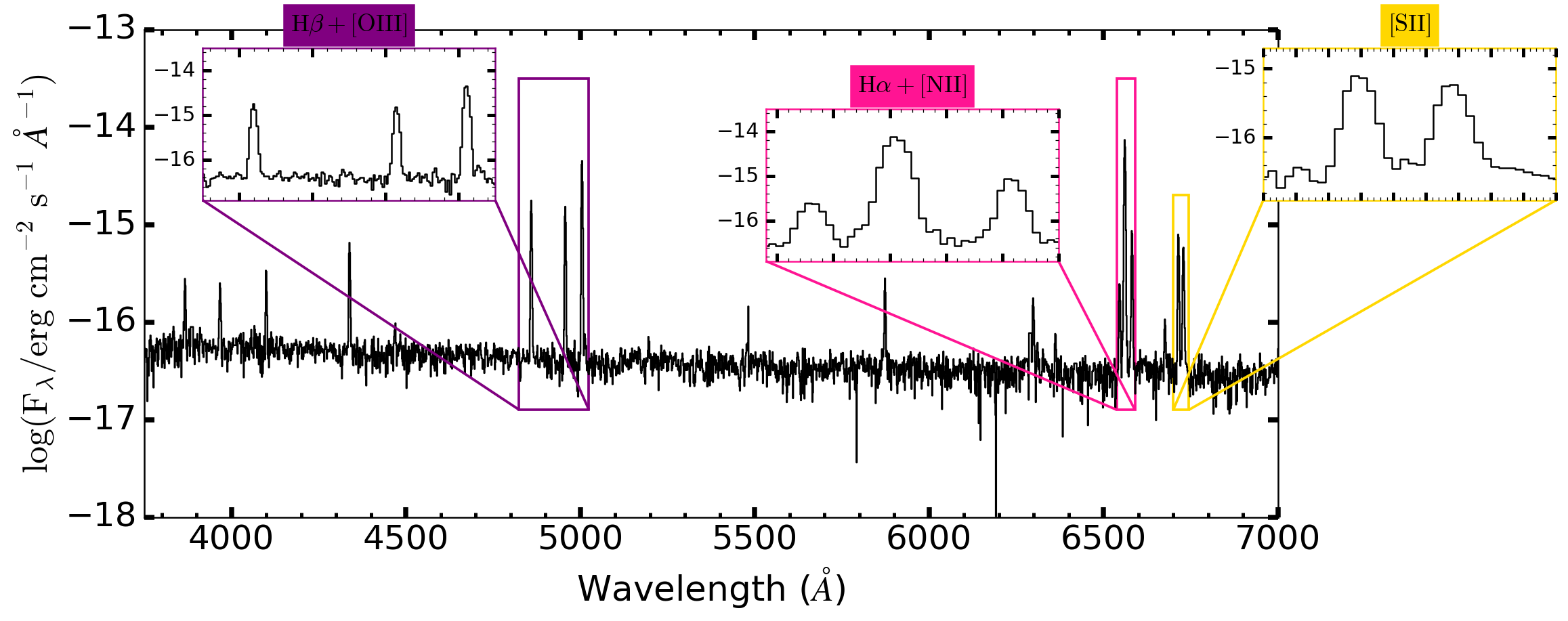}
    \caption{A typical MMT/Binospec spectrum of a \HII region in \object{NGC 99}. The flux is shown in log scale to show the level of the underlying stellar continuum. The three insets highlight the emission lines of interest in this study.}
    \label{fig:spectra}
\end{figure*}

Each mask was observed for a total of 2400 seconds using 4 exposures of 600 seconds each during the same night. We used the 270 lines mm$^{-1}$ grating of Binospec with a central wavelength of 5800 \AA$~$for each slit at a dispersion of 1.3 \AA$~\text{pix}^{-1}$. This allowed us to cover the entire wavelength range of emission lines observed from 3900\AA$~$- 9240\AA$~$in one setting. The data were reduced using the Binospec Data Reduction Pipeline, which carried out the bias correction, flat-fielding, wavelength calibration, relative flux correction, co-addition, and extraction to 1D spectra \citep{kansky_2019PASP..131g5005K}. Of the 31 regions observed, the spectra of 26 regions had measurable emission lines. The slit locations can be seen in Figure \ref{fig:HII_regions}. Figure \ref{fig:spectra}, a spectrum of region 16, shows a representative \HII region spectrum produced by the Binospec instrument with the lines we aim to measure.

After correcting the spectra for redshift, emission lines were found using the $\texttt{find\_lines\_derivative}$ method from the \texttt{specutils} python package \citep{specutils} and identified by matching the line 

\begin{deluxetable}{lC}[h!]
\tablenum{1}
\label{tab:galaxy properties}
\tablecaption{Properties of NGC 99}
\tablewidth{0pt}
\tabletypesize{\small}
\tablehead{
\colhead{Parameter} & \colhead{Value}
}
\startdata
{\rm R.A.}\tablenotemark{a}  ($\alpha_{2000}$) & 00^{\text{h}}23^{\text{m}}59^{\text{s}}.422\\
{\rm Decl.}\tablenotemark{a}  ($\delta_{2000}$) & +15\degree46^{\prime}13^{\prime\prime}.04\\
{\rm Morphological~Type}\tablenotemark{b} & $\rm SABc$\\
{\rm Inclination}\tablenotemark{c}& $\rm 20^{\circ}$\\
{\rm Redshift}\tablenotemark{d} & 0.01771 \\
{\rm Distance}\tablenotemark{e}& 79.4~$\rm Mpc$\\
{\rm Stellar~mass ($\rm M_{\star,global}$)}\tablenotemark{c} & $4.17\times10^{10}~\rm M_\odot$ \\
$\rm R_{25}$\tablenotemark{c} & 11.46~$\rm kpc$ \\
$R_{e}$\tablenotemark{c} & 8.51~$\rm kpc$ \\
{\rm E(B-V)}\tablenotemark{f} & 0.0481\\
{\rm SFR$\rm _{global}$}\tablenotemark{g} & 2.45~$\rm M_\odot~yr^{-1}$\\
$R_{\rm HI}$\tablenotemark{h} & 45.8~$\rm kpc$ \\
{\rm \HI mass}\tablenotemark{h} & $1.68\times10^{10}~\rm M_\odot$\\
{\rm Circular~velocity}\tablenotemark{h} & 292~$\rm km~s^{-1}$\\
{\rm Gas~dispersion}\tablenotemark{i} & 14~$\rm km~s^{-1}$\\
{\rm Halo~mass}\tablenotemark{j} &  $5.0\times10^{11}~\rm M_\odot $\\
{\rm Virial~radius}\tablenotemark{j} & 207~$\rm kpc$
\enddata

\tablenotetext{a}{\cite{2MASS_2006AJ....131.1163S}}
\tablenotetext{b}{\cite{Buta_2019MNRAS.488..590B}}
\tablenotetext{c}{\cite{Padave_2024ApJ...960...24P}}
\tablenotetext{d}{\cite{Springob_2005ApJS..160..149S}}
\tablenotetext{e}{Calculated using the distance modulus value from \cite{2012AA...546A...2S}.}
\tablenotetext{f}{\cite{Schlafly_2011ApJ...737..103S}}
\tablenotetext{g}{Estimated from FUV and 22$\mu$m data adopted from \cite{Padave_2024ApJ...960...24P}.}
\tablenotetext{h}{Based on VLA D-conf. observations (Gim et al., in prep.)}
\tablenotetext{i}{Based on the peak of the distribution of velocity dispersions in the VLA D-configuration \HI image. We adopt the peak value because some of the values at the central regions are impacted by beam smearing of the steeply varying rotation curve.}
\tablenotetext{j}{Halo mass and virial radius were estimated from the stellar mass using prescriptions by \citep{Kravtsov_2018AstL...44....8K} and applying modifications based on the findings of \cite{Mandelbaum_2016MNRAS.457.3200M}. The full sample will be described in Borthakur et al. (in prep).}
\end{deluxetable}

center with an emission line database. The local continuum level within a $\sim100$\AA\, window was measured using \texttt{specutils}' $\texttt{fit\_generic\_continuum}$ function and subtracted from each emission line. The total integrated flux of each line was then calculated using the amplitude and standard deviation of a modeled Gaussian. Using pyFIT3D \citep{Lacerda_2022NewA...9701895L}, we modeled the stellar spectrum of our HII region with the strongest continuum level to investigate the effect of stellar absorption, which we found to be minimal. Therefore, for other targets where the continuum is weaker, the stellar absorption is not modeled as it would not be significant compared to the strength of the emission lines.

Our absolute flux calibration has some uncertainty. However, this should have no impact on our metallicity measurements and very little on the Balmer decrement. To ensure an accurate SFR estimate, we cross-calibrated our ${\rm H}\alpha$ fluxes with VATT narrow-band ${\rm H}\alpha$ imaging \citep{Padave_2024arXiv240716690P}.
The ${\rm H}\alpha$ flux of each region is reported in Table \ref{tab:indicators}. We investigated the shape of the ${\rm H}\alpha$ emission line, which is the strongest line in these spectra, to look for evidence of multiple component emission. We found that all but one of the emission lines fit well with a single Gaussian profile. The spectrum from Region 2 shows a slight wing towards shorter wavelengths.

\subsection{Optical g- and r-band data}\label{subsec:VATTdata}

\indent Optical continuum g- and r-band imaging of \object{NGC 99} was obtained on UT 2020 October 07 using the VATT4k CCD imager at the 1.8 m Vatican Advanced Technology Telescope (VATT) operated by the Mt. Graham Observatory. The total exposure times in g and r are 600 and 1200 seconds, respectively. The observation setup and data reduction are described in \cite{Padave_2024ApJ...960...24P}.

We utilize the r-band image and \texttt{statmorph} \citep{Rodriguez-Gomez_2019MNRAS.483.4140R} to estimate position angle, ellipticity, and inclination. We then use these parameters and the right ascension and declination of each region to calculate the \HII region's semimajor axis value. We assume the galaxy to be circular in projection, thus taking the semimajor value to be the \HII region's galactocentric radius.

\subsection{VLA \HI 21cm Data}\label{subsec:HIdata}
The \HI 21\,cm emission from NGC 99 was observed with the Karl G. Jansky Very Large Array (VLA) in the D-configuration as a part of VLA-DIISC project (Gim et al. in prep.). The observations were performed from November 9 through November 16, 2019 (program ID: 19B-183) for a total of 6.5 hours with a channel spacing of 5.208~kHz within the bandwidth of 16~MHz at the central frequency of 1395~MHz. The data reduction was carried out with the Common Astronomy Software Application (CASA; \citealt{McMullin_2007ASPC..376..127M}) version 5.6.1 according to the general reduction schemes for \HI spectroscopy. The absolute flux density scale and bandpass were calibrated using 3C48, while the complex gain was calibrated using J2340+1333. Hanning smoothing was applied to the data to mitigate Gibbs ringing phenomena, resulting in an effective velocity resolution of 2.2 \kms, doubling the original velocity width.

The image cube was made by the CASA task $tclean$ with a pixel size of 6\arcsec, the Briggs weighting function, and the robust value of 0.5 for the optimal sensitivity and synthesized beamsize. The cleaning was performed until the maximum residual reached 1.5$\sigma$, where $\sigma$ is the root-mean-squared (RMS) noise in the image. The image cube was spatially smoothed to the synthesized beamsize of 50.5\arcsec$\times$47.5\arcsec over all the channels. The final image cube has a sensitivity of 0.69 mJy $\rm beam^{-1}$ (2.2 \kms)$^{-1}$ corresponding to a column density of $6.81 \times 10^{18}$~cm$^{-2}$.

The \HI 21\,cm emission from NGC\,99 was recovered by the Source Finding Application-2 (SoFiA-2, \cite{Serra_2015MNRAS.448.1922S}; \cite{Westmeier_2021ascl.soft09005W}) above 3$\sigma$, where it was found within the velocity range of 5125.4 -- 5312.3 km s$^{-1}$. The velocity width of the \HI spectrum is $\rm W_{50} = 130.54$~and $\rm W_{20}=154.78$~km s$^{-1}$ at 50\% and 20\% of the maximum intensity, respectively. The \HI mass was estimated to be $\rm M_{HI} = (1.68 \pm 0.03) \times 10^{10}$~M$_{\odot}$ assuming a luminosity distance of 79.4~Mpc.

\subsection{COS Spectra} \label{sec:cos}
The ultraviolet spectrum of the quasi-stellar object (QSO) SDSS J002330.58+154744.9, which is at an impact parameter of 159 kpc from the center of NGC 99, was obtained using the G130M medium resolution grating of the Cosmic Origin Spectrograph (COS;\citealt{Osterman_2011Ap&SS.335..257O,Green_2012ApJ...744...60G}) aboard the Hubble Space Telescope under observing program GO-14071 (PI: Borthakur).
After processing with the standard COS pipeline, the multiple G130M spectra were co-added and binned by 3 pixels, resulting in a spectral bin size of $\sim$7\:km\:s$^{-1}$ in the observed wavelength coverage of $\lambda=$ 1152 -- 1453 \AA. All absorption features associated with the QSO, the Milky Way's ISM, and NGC\,99's CGM were identified through visual inspection. We detected \ion{H}{1} $\lambda1215$ (Ly$\alpha$) in absorption in the CGM of \object{NGC 99}. 

To fit the \Lya profile, we determined the continuum bracketing the absorption by using feature-free regions within $\pm$1000\:km\:s$^{-1}$. The continuum was estimated using a Legendre polynomial of order 2 using the procedure of \cite{sem04}, which was then used to produce the normalized spectrum. We fit Voigt profiles to each feature using the software of \cite{fitz97}, and techniques similar to \cite{tum13}. These fits derived the velocity centroids, Doppler \textit{b}-values, and column densities for the profile, which are listed in Table \ref{tab:lya}. The measurement uncertainties were derived using the error analysis methods of \cite{sem92}.

\begin{figure}[h!]
    \centering
    \includegraphics[scale=0.35]{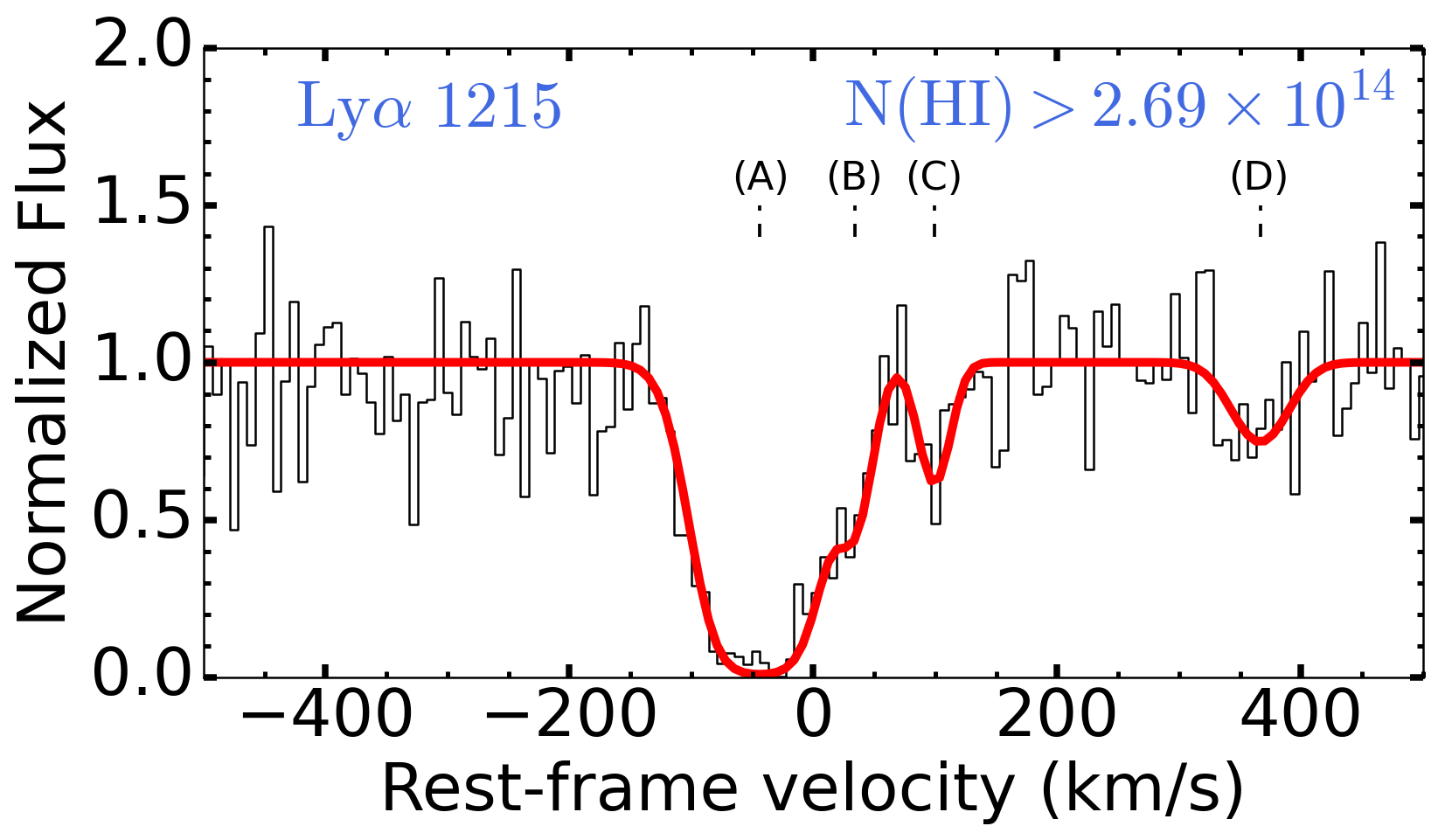}
    \caption{COS spectra of the \Lya transition plotted at the rest-frame of \ngc ($z=0.0177$). The best combined profile is plotted in red, and the resulting column density estimate is shown in the upper right. The dominant absorption component (A) is -44~km~s$^{-1}$ offset from the systemic velocity of the galaxy with three weaker components (B-D) located at 34~km~s$^{-1}$, 99~km~s$^{-1}$, and 366~km~s$^{-1}$. \Lya is the only detected species associated with the galaxy's CGM. Measurements for \Lya and upper limits for the metal lines are presented in Table~\ref{tbl:CGM}.}
    \label{fig:lya}
\end{figure}

We find four components (labeled A-D) of \Lya and fit each separately. Due to saturation, the best fit Voigt profile for the dominant component (A) gives a lower limit on the column density of $N$(\ion{H}{1}$)>2.69\times10^{14}$ cm$^{-2}$ (Figure \ref{fig:lya}). The three weaker components (B-D) have column density estimates of $N$(\ion{H}{1}$)=1.67\times10^{13}$, $1.13\times10^{13}$, and $1.20\times10^{13}$ cm$^{-2}$. The centroid for the dominant component is -44~km~s$^{-1}$ offset from the systemic velocity of \object{NGC 99} while the weaker components are located at 34~km~s$^{-1}$, 99~km~s$^{-1}$, and 366~km~s$^{-1}$. 

\begin{deluxetable*}{lcccccr}
\tablenum{2}
\label{tbl:CGM}
	\tablecaption{Measurements from QSO Absorption Spectroscopy  Tracing the CGM of \ngc \label{tab:lya}}
	\tablecolumns{6}
	\tablehead{
	\colhead{Species} & \colhead{$\lambda_{\rm rest}$} & \colhead{$\rm W_{rest}$\tablenotemark{a}} & \colhead{Component} & \colhead{Centroid\tablenotemark{b}} & \colhead{Doppler $b$} & \colhead{log $N$ } \\
	\colhead{}	 & \colhead{(\AA)} & \colhead{(m\AA)} & \colhead{Label} & \colhead{(km\:s$^{-1}$)} & \colhead{(km\:s$^{-1}$)} & \colhead{(log\:cm$^{-2}$)}
	}
	\startdata
	\ion{H}{1} & 1215 & 513$\pm$52 & (A) &-44$\pm$8 & 42$\pm$0.1 & $\geq$14.4$\pm$0.2\\
    &&&(B)&34$\pm$14&17$\pm$42&13.2$\pm$0.6\\
    &&&(C)&99$\pm$11& 15$\pm$41 & 13.1$\pm$0.5\\
    &&&(D)&366$\pm$21 & 30$\pm$1 & 13.1$\pm$0.4\\
\ion{Si}{2} & 1302 & $\le$ 119\tablenotemark{c} & -- & -- & -- & $\leq$12.86\\
\ion{Si}{3} & 1206 & $\le$ 144\tablenotemark{c} & -- & -- & -- & $\leq$12.83\\
\ion{Si}{4} & 1393 & $\le$ 144\tablenotemark{c} & -- & -- & -- & $\leq$13.21\\
\ion{C}{2} & 1334  & $\le$ 143\tablenotemark{c} & -- & -- & -- & $\leq$13.85\\
\ion{N}{5} & 1238  & $\le$ 124\tablenotemark{c} & -- & -- & -- & $\leq$13.77\\
	\enddata
	\tablenotetext{a}{Equivalent widths as estimated directly from the data. The error is calculated from both continuum fitting and statistical errors.}
	\tablenotetext{b}{Velocity centroid values are reported with respect to the rest-frame of NGC~99 ($z= 0.0177$).}
 \tablenotetext{c}{The values represent 3$\sigma$ upper limits.}
\end{deluxetable*} 

\section{Results} \label{sec:results}

\subsection{Dust Extinction} 

\begin{figure*}[h!]
\begin{minipage}{0.49\textwidth}
    \centering
    \includegraphics[width=.93\linewidth]{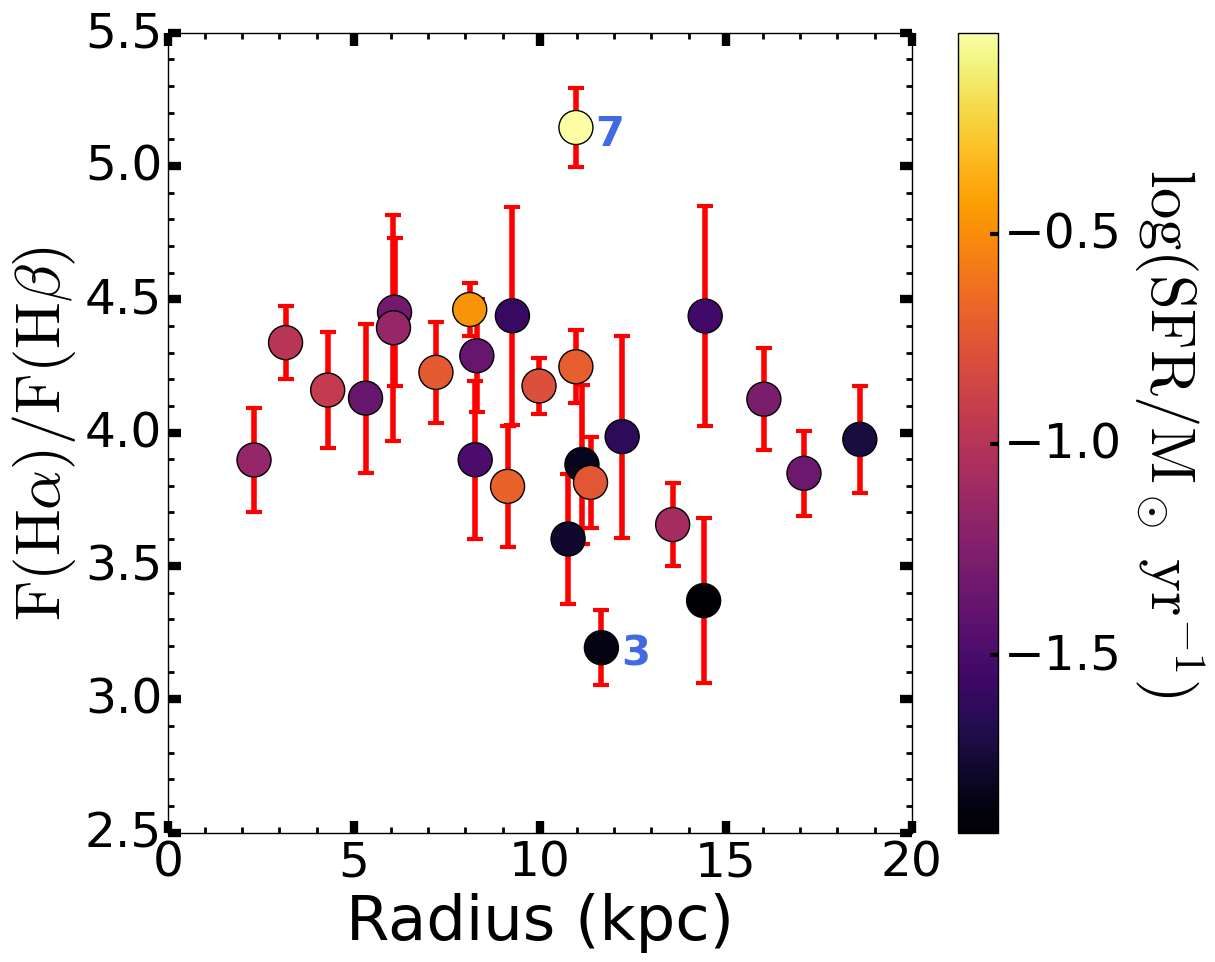}
\end{minipage}%
\begin{minipage}{0.49\textwidth}
    \centering
    \includegraphics[width=.90\linewidth]{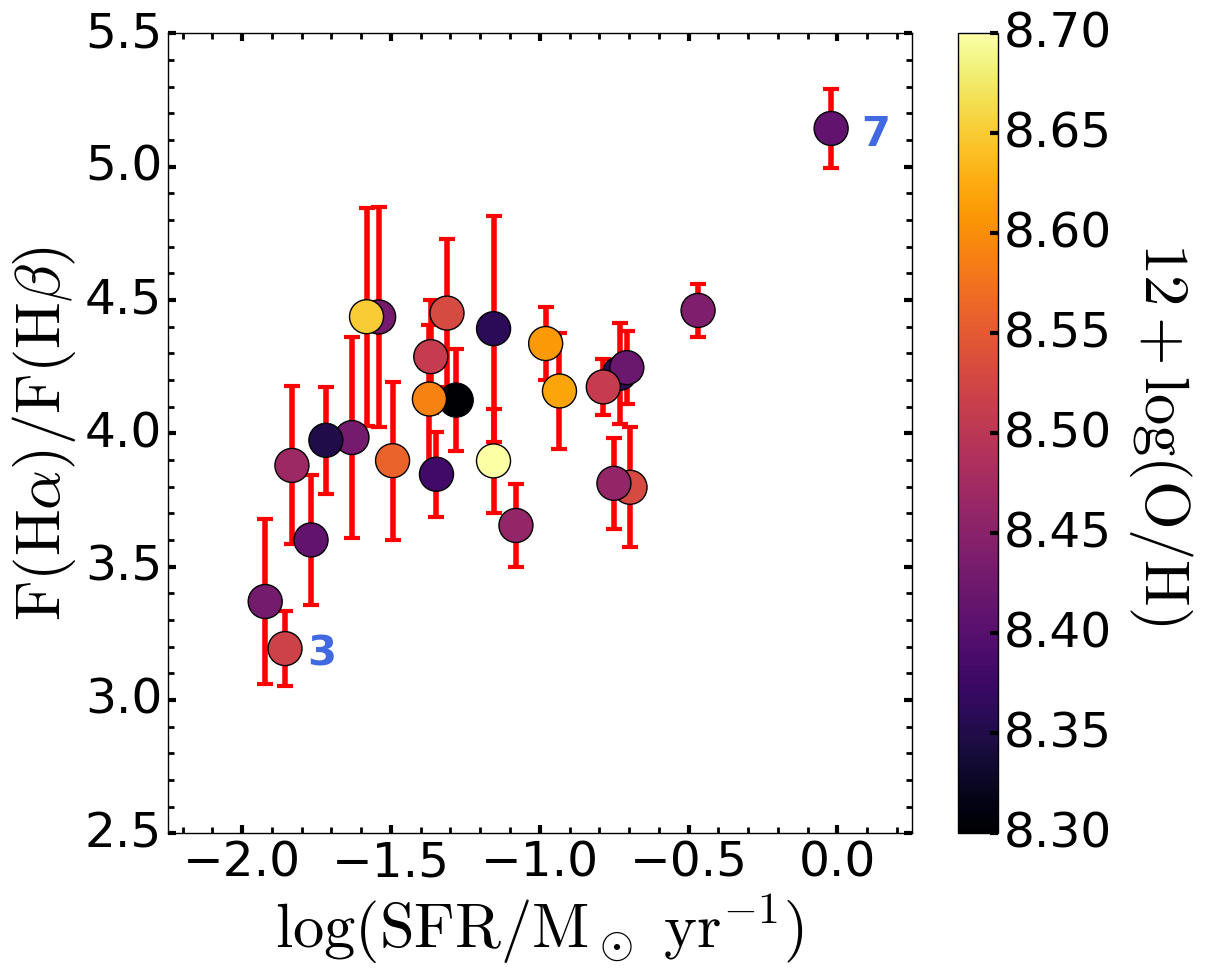}
\end{minipage}
\caption{The Balmer decrement of each \HII region as a function of its galactocentric radius (left) and star-formation rate (right). The associated error bars are overplotted in red. The observed \HII regions within 10~kpc of the center of NGC 99 show a flat radial distribution. \HII regions outside of 10 kpc show wide scatter. \label{fig:balmer}}
\end{figure*}

We investigated the dust content of the \HII regions using the Balmer decrement, F(H$\alpha$)/F(H$\beta$) flux, across the galaxy. The Balmer decrement of each region is listed in Table \ref{tab:indicators}. The left panel of Figure \ref{fig:balmer} shows the Balmer decrement for the 26~\HII regions as a function of galactocentric radius. All the points show F(H$\alpha$)/F(H$\beta$) values greater than the expected intrinsic value of 2.86 from Case B recombination in the absence of dust. Within 10~kpc of the center, the regions show a flat Balmer decrement, while at radii greater than 10~kpc, the points show a wide scatter. One point in particular, region 7 at 11~kpc, shows a high F(H$\alpha$)/F(H$\beta$) value of $\approx 5.14$ while the lowest value of 3.19 was observed in region 3 at 11.7 kpc.

The right panel of Figure \ref{fig:balmer} reveals a connection between the Balmer decrement and the SFR of the \HII regions where a higher SFR is associated with higher Balmer ratios. From the Schmidt-Kennicutt (SK) law, we expect higher SFRs with larger cold gas surface densities, $\Sigma_{{\rm gas}}$ \citep{Schmidt_1959ApJ...129..243S, Kennicutt_1998ApJ...498..541K}. For a fixed dust-to-mass ratio, the dust column density and the gas column densities are coupled; the observed correlation is not surprising.

Dust has also been used as a proxy for gas metallicity; however, the color-coded points of the right panel of Figure \ref{fig:balmer} reveal that the Balmer decrement of the \HII regions has little dependence on the gas-phase metallicity of the regions. Therefore, we conclude that metallicity is not the primary factor affecting the trend between the dust content and the SFR.

\subsection{Gas-phase Metallicity}\label{sec:metallicity}

\begin{figure*}[h!]
\begin{minipage}{0.49\textwidth}
\centering
 \includegraphics[width=.93\linewidth]{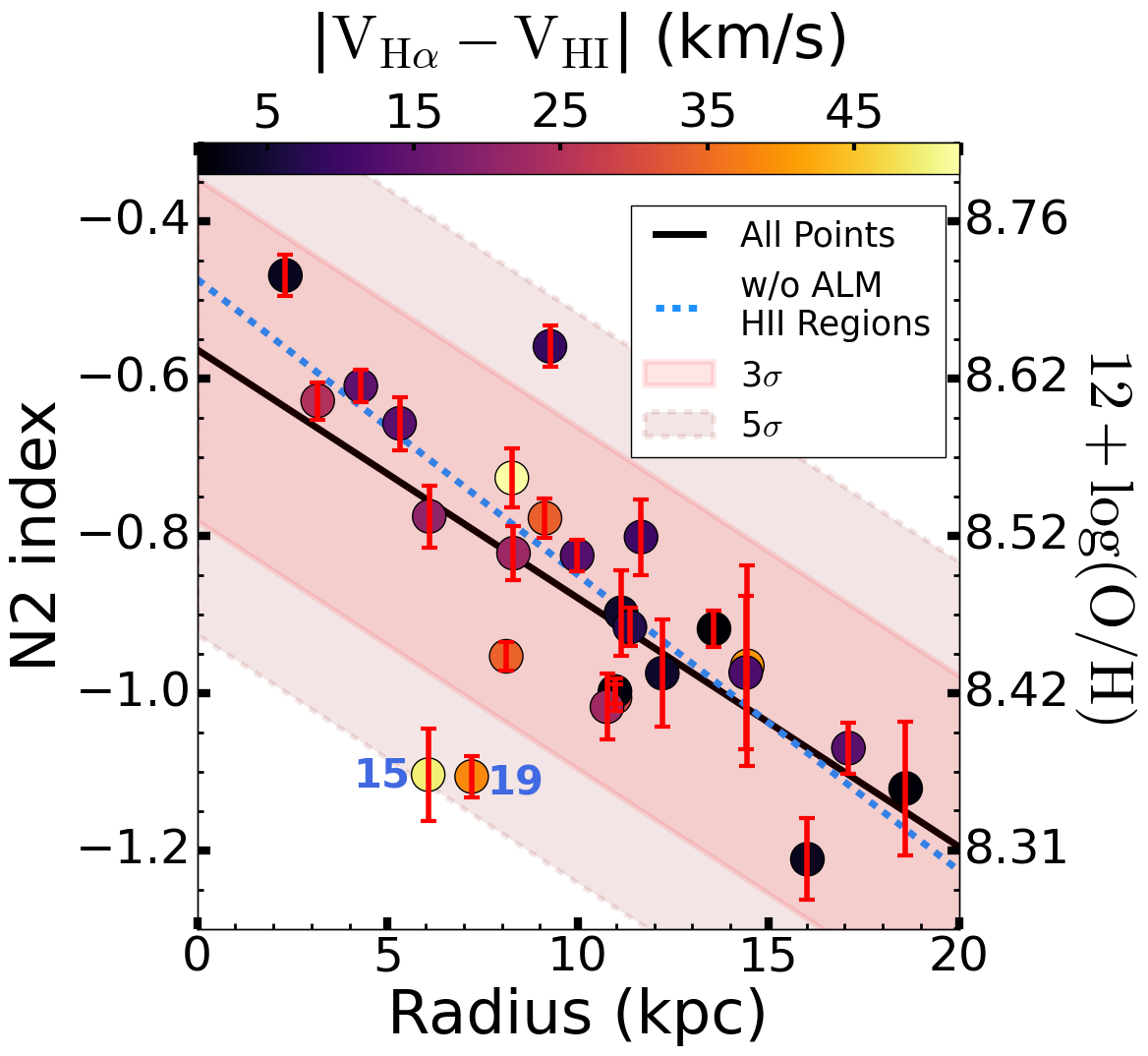}
\end{minipage}%
\begin{minipage}{0.49\textwidth}
\centering
 \includegraphics[width=.90\linewidth]{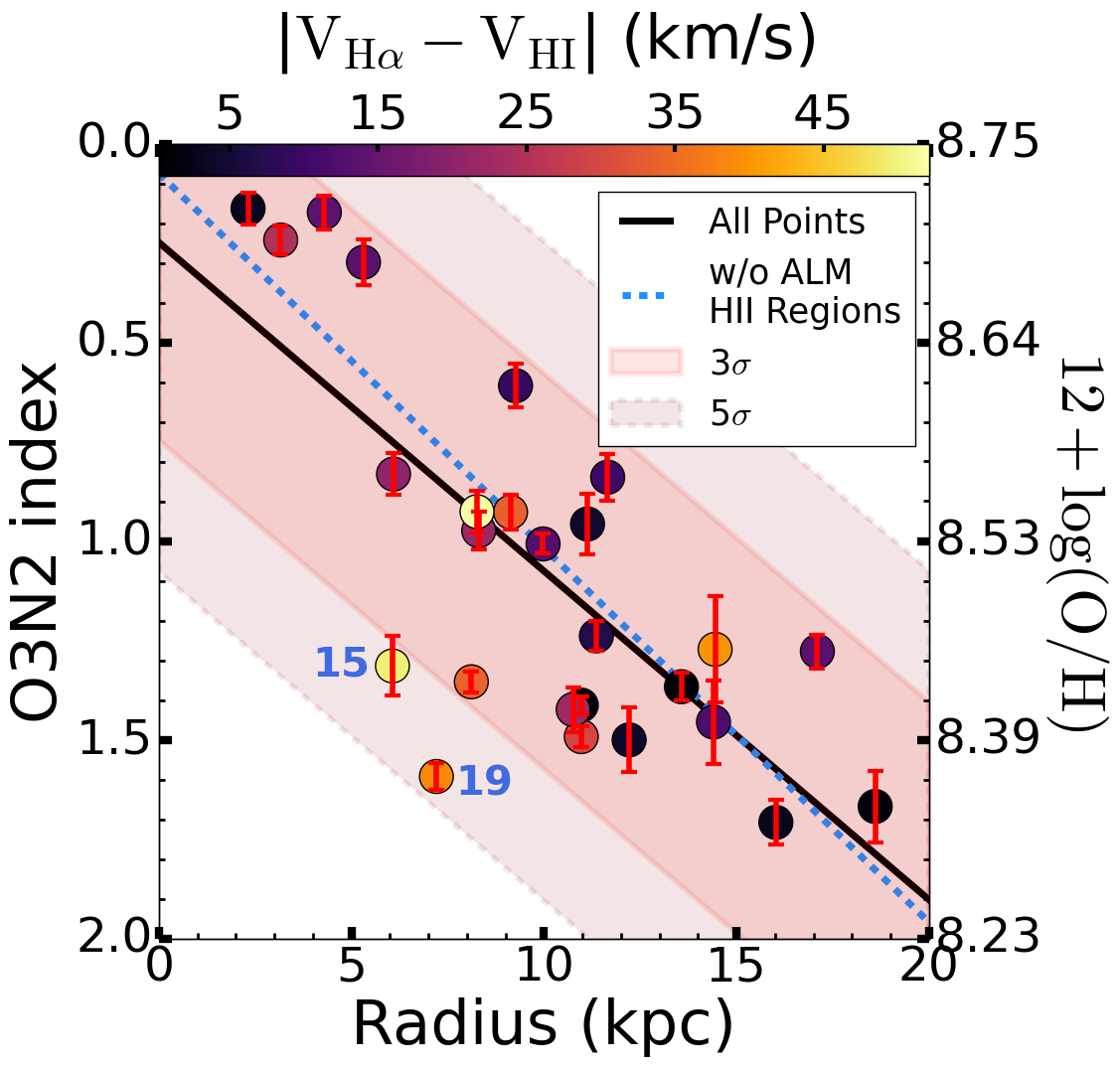}
\end{minipage}
\caption{The metallicity radial gradient from both the N2 (left) and the O3N2 (right) indices. The right-side y-axis in both panels shows the corresponding gas-phase metallicity calculated from the calibration equations derived in \cite{Curti_FMR_2020MNRAS.491..944C}. The black solid line shows the best linear fit to all of the data, and the red and dark red shaded regions signify the $3\sigma$ and $5\sigma$, respectively, confidence intervals of the fit. For y = 12+log(O/H), the slopes of the fit are -0.017 and -0.020 dex/kpc, respectively. The colorbar shows the absolute difference between the ${\rm{H}}\alpha$ emission line velocity offset and the mass-weighted \HI 21\,cm velocity of the ISM in the region. Regions 15 and 19, which are labeled and are at 6.1 and 7.2 kpc, show both anomalously low metallicity and a high velocity difference. The blue line shows the fit without the two ALM regions. The fits (slopes and intercepts) are presented in Table \ref{tab:linearfits}.}
\label{fig:N2-O3N2_vs_R}
\end{figure*}

\begin{figure*}[h!]
\plottwo{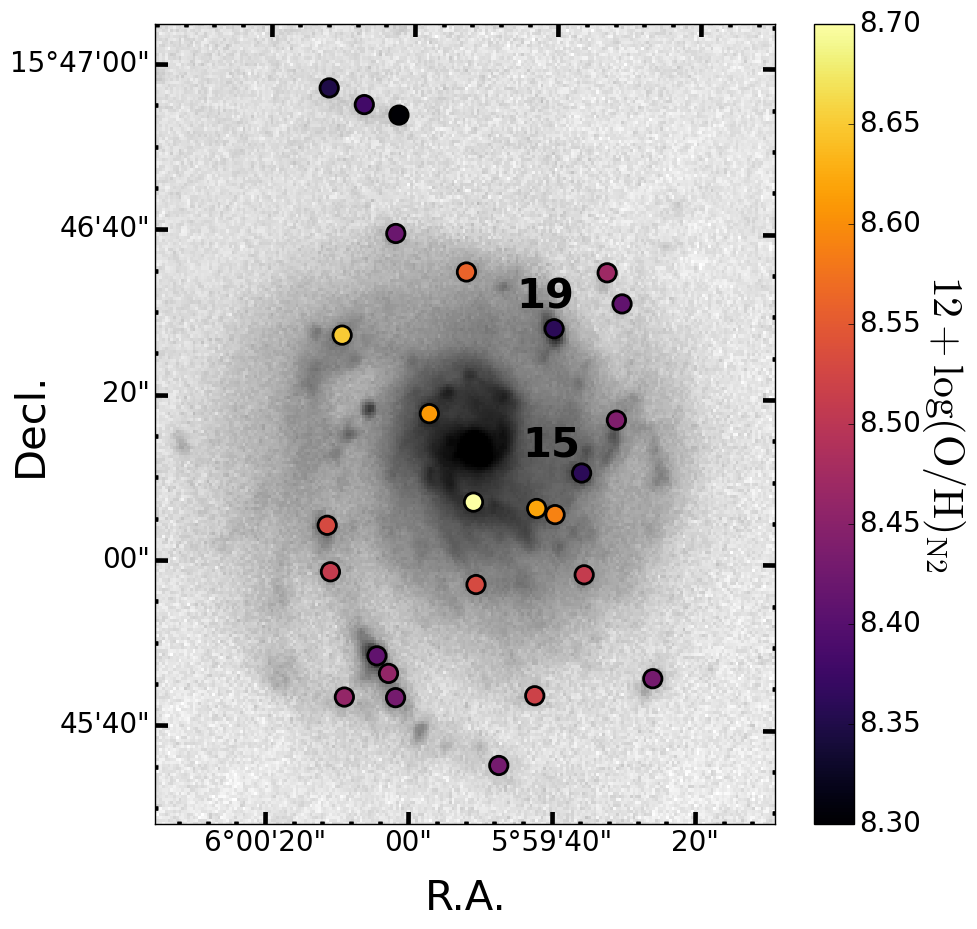}{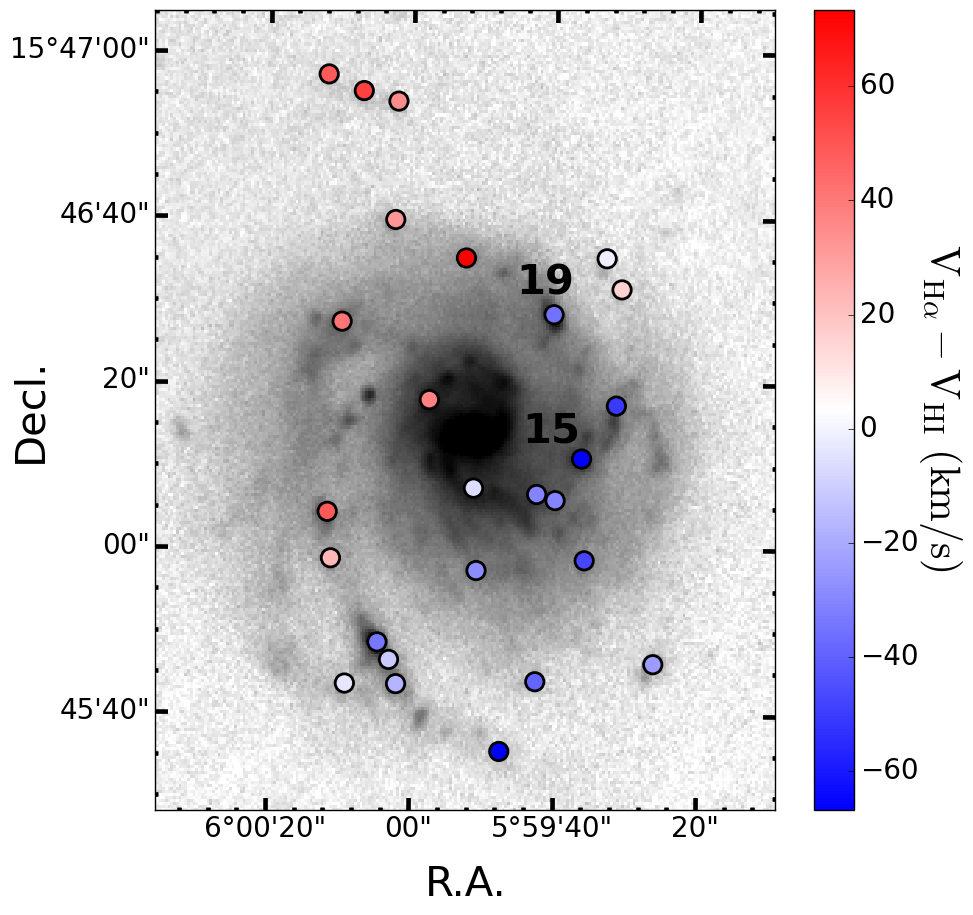}
\caption{r-band images of the NGC 99 with the locations \HII regions indicated by the circles, which are color-coded based on their gas-phase metallicity (left) and velocity offset of the ${\rm{H}}\alpha$ emission line (right). Regions 15 and 19 show abnormally low metallicities and high velocities relative to their surrounding regions.}
\label{fig:maps}
\end{figure*}

We investigate how the metallicity of the \HII regions in the galaxy varies as a function of galactocentric radius. We measured the flux of the strong emission lines ${\rm{H}}\alpha$, ${\rm{H}}\beta$, ${\rm{[OIII]}}\lambda5007$, ${\rm{[NII]}}\lambda6583$, and ${\rm{[SII]}}\lambda6717, 31$. The ratios of these fluxes provide metallicity indicators N2 and O3N2, which are defined as

\begin{equation}\label{equ:N2}
    {\rm{N2}} = {\rm{log}}\left(\frac{{\rm{[NII]}}\lambda6583}{{\rm{H}}\alpha}\right)
\end{equation}

\begin{equation}\label{equ:O3N2}
    {\rm{O3N2}} = {\rm{log}}\left(\frac{{\rm{[OIII]}}\lambda5007}{{\rm{H}}\beta}\frac{{\rm{H}}\alpha}{{\rm{[NII]}}\lambda6583}\right)
\end{equation}

\vspace{0.3cm}

where the lines refer to line fluxes \citep{Searle_1971ApJ...168..327S,Alloin_1979A&A....78..200A,Denicolo_2002MNRAS.330...69D,Kewley_2002ApJS..142...35K,Pettini_2004MNRAS.348L..59P,Nagao_2006A&A...459...85N,Marino_2013A&A...559A.114M,Doptia_2016Ap&SS.361...61D,Curti_2017MNRAS.465.1384C,Bian_2018ApJ...859..175B,Maiolino_2019A&ARv..27....3M,Curti_FMR_2020MNRAS.491..944C}. The N2 and O3N2 values for each region are given in Table \ref{tab:indicators}. Both the indicators have the benefit of being insensitive to dust attenuation since the two lines in each line ratio (H$\alpha$/[\ion{N}{2}]$\lambda6583$ and  [\ion{O}{3}]$\lambda5007$/H$\beta$) are close in wavelength space ($\sim$20\AA). The indicators serve well as proxies for the metallicity of the systems and can be converted into oxygen gas-phase metallicities $12 + \rm log(O/H)$. We follow the prescription from \cite{Curti_FMR_2020MNRAS.491..944C} to obtain gas-phase metallicities.

Figure \ref{fig:N2-O3N2_vs_R} shows N2 and O3N2 as a function of galactocentric radius with the corresponding gas-phase metallicity on the right y-axis. The black solid line shows the best linear fit between the galactocentric radius and gas-phase metallicity, while uncertainty is given by the shaded regions. The slope, intercept, and the associated errors were estimated using \texttt{polyfit} from NumPy. When fitting for 12 + log(O/H) instead of N2 and O3N2 as a function of galactocentric radius, the radial gradients are -0.017 and -0.020 dex kpc$^{-1}$, respectively. Further information can be found in Table \ref{tab:linearfits}. The filled circles are color-coded by $\lvert{\rm V}_{{\rm H}\alpha}-{\rm V}_{{\rm HI}}\rvert$, the absolute difference between the centroid of the ${\rm H}\alpha$ emission line and the mass-weighted \HI 21\,cm velocity of each region. Both indicators show the expected trends of decreasing metallicity with increasing radius in general, with N2 and 
O3N2 having a Spearman-$\rho$ value of -0.60 and -0.63, respectively.

Interestingly, two regions at radii 6.1 and 7.2 kpc (labeled 15 and 19 in Figure \ref{fig:HII_regions}, respectively) show an abnormal drop in metallicity compared to others at similar radii, regardless of which metallicity indicator is used. These regions are more than $3\sigma$ below the linear fits shown in Figure \ref{fig:N2-O3N2_vs_R}. In addition, these regions also show a high velocity offset between ${\rm V}_{{\rm H}\alpha}$ and ${\rm V}_{{\rm HI}}$. Both trace the spiral arms and, on visual inspection, do not show any morphological differences. 

We performed a Cook's distance statistical outlier test and confirmed that these regions are outliers in the data. Masking the two anomalously low metallicity (ALM) \HII regions, we refit the data and found that the radial gradient for both N2 and O3N2 became steeper by 0.03 dex. The new gradients are shown by the blue dashed line in Figure \ref{fig:N2-O3N2_vs_R}.

Figure \ref{fig:maps} shows the spatial locations of the \HII regions with filled circles color-coded based on their metallicity and velocity offset of the ${\rm H}\alpha$ emission line from the ISM kinematics. Region 15 lies in one of the spiral arms of \object{NGC 99} and is surrounded by points with higher metallicity while also having a higher velocity offset. Region 19 shows a negative velocity offset despite its colocated ISM showing a positive velocity consistent with rotation. Combining the metallicity and velocity information presented in Figure \ref{fig:N2-O3N2_vs_R}, we suggest that the star formation in these regions is fueled by low metallicity gas that has been accreted into the disk of the galaxy.

Alternative explanations for the ALM \HII regions may be azimuthal and arm-interarm variations in the oxygen abundance and deviations from the general radial profiles, which have been observed and described
in many galaxies \citep{KennicuttGarnett_1996ApJ...456..504K, MartinBelley_1996ApJ...468..598M, CedresCepa_2002A&A...391..809C, RosalesOrtega_2011MNRAS.415.2439R, Cedres_2012A&A...545A..43C, Li_2013ApJ...766...17L, Sanchez_2015A&A...573A.105S, Sanchez-Men_radial_inflow_2016ApJ...830L..40S, SanchezMenguiano_2017A&A...603A.113S, SanchezMenguiano_2019ApJ...882....9S}. It is also possible that these \HII regions are extraplanar and look like part of the disk due to projection effects. However, that is not likely in this case as NGC 99 is almost face-on (\textit{i}=20$^{\circ}$), so projection effects are minimal. Therefore, we conclude that the ALM \HII regions are due to the accretion of low metallicity gas. Similar conclusions were also derived by other recent studies such as \cite{Howk_a_2018ApJ...856..166H}, \cite{Hwang_2019ApJ...872..144H}, and \cite{Scholz_2021MNRAS.505.4655S}.

\subsection{[SII] Deficiency}\label{sec:SII de}

\begin{figure*}
\begin{minipage}{0.49\textwidth}
     \centering
     \includegraphics[width=.99\linewidth]{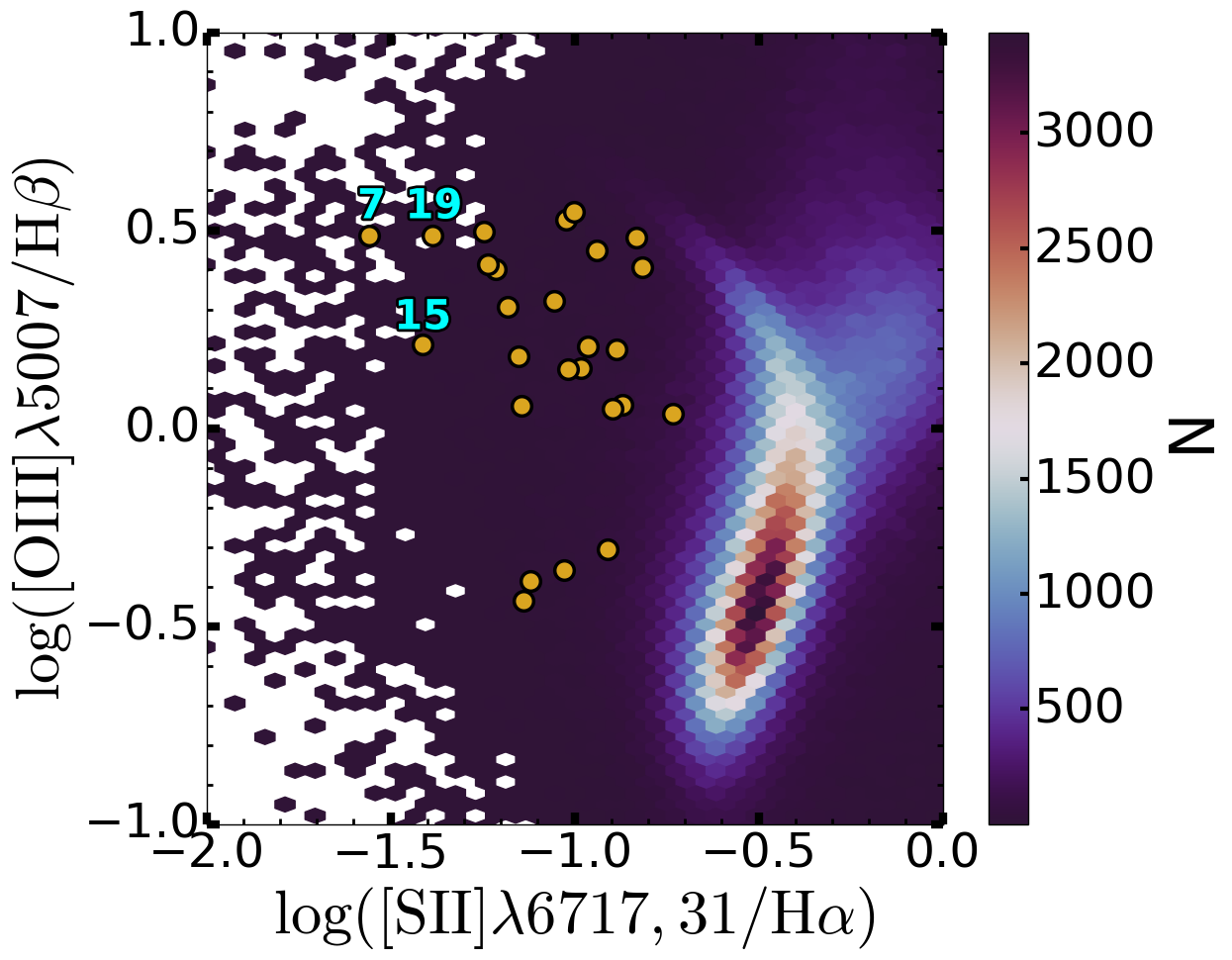}
\end{minipage}%
\begin{minipage}{0.49\textwidth}
     \centering
     \includegraphics[width=.98\linewidth]{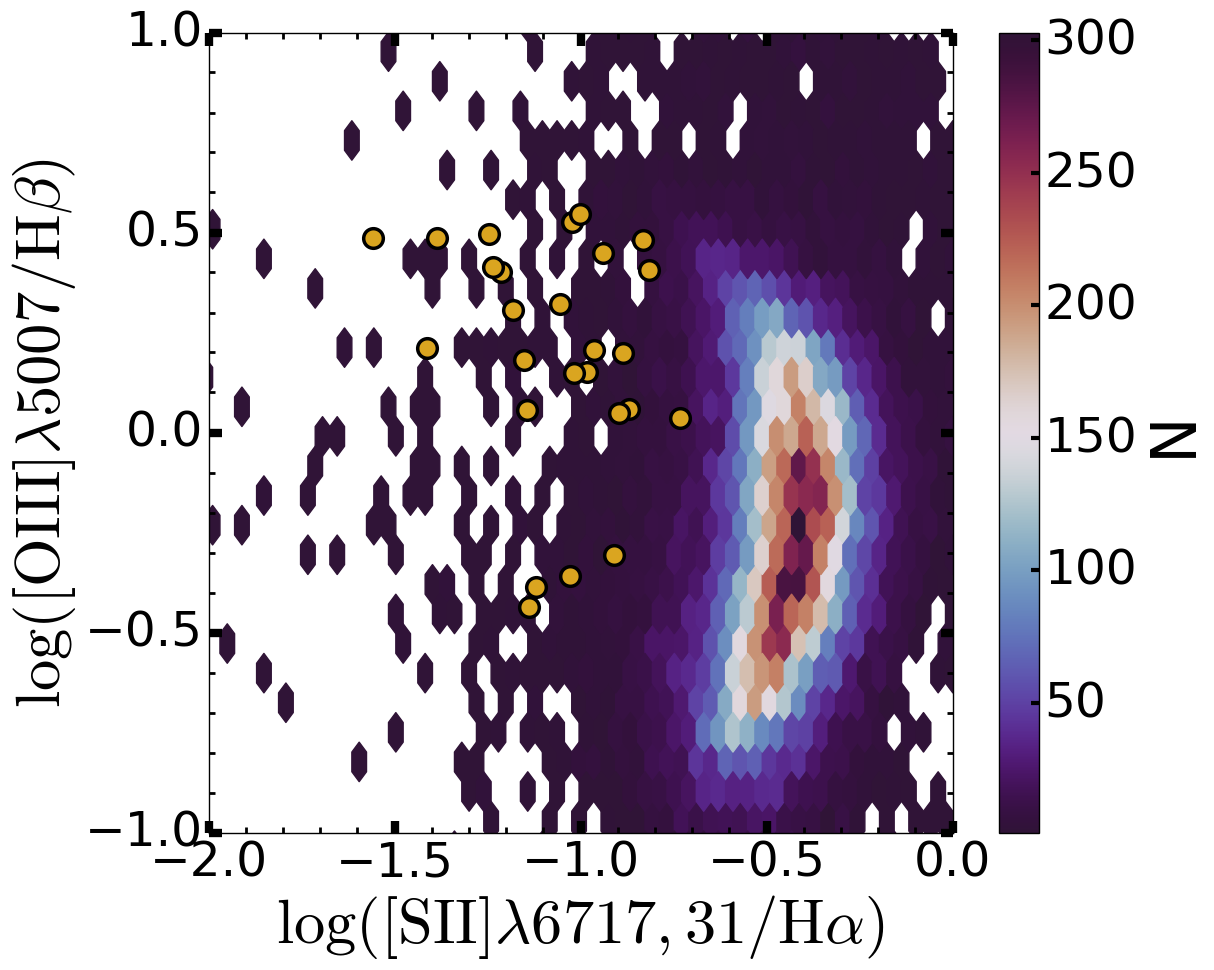}
\end{minipage}
    \caption{Distribution of 26 \HII regions (filled gold circles) on a background of hexagon bins from SDSS-DR8 galaxies (left) and from CALIFA HII regions (right). The colorbar shows the number of galaxies or HII regions in each bin. The data points show a 0.5-1.0 dex offset in [SII] from the majority of SDSS galaxies and CALIFA HII regions.}
    \label{fig:BPT}
\end{figure*}

We make measurements of the [SII]$\lambda6717,31$ emissions lines and measure [SII]-based metallicity indicators such as S2, N2S2H$\alpha$, and RS32 \citep{Denicolo_2002MNRAS.330...69D,Yin_2007A&A...462..535Y, Doptia_2016Ap&SS.361...61D,Curti_2017MNRAS.465.1384C,Maiolino_2019A&ARv..27....3M,Curti_FMR_2020MNRAS.491..944C}. For example, the S2 index, defined as

\begin{equation}\label{equ:S2}
    {\rm{S2}} = {\rm{log}}\left({\frac{{\rm{[SII]}}\lambda6717,31}{{\rm{H}}\alpha}}\right),
\end{equation}

\noindent shows no radial dependence with galactocentric radius. Overall, the [SII]-based indicators show large variation compared to N2 or O3N2. 

To investigate it further, we constructed a line ratio diagram (Figure \ref{fig:BPT}) using the line ratios of ${\rm{[OIII]/H}}\beta$ and ${\rm{[SII]/H}}\alpha$ from our 26 \HII regions. The left panel of Figure \ref{fig:BPT} shows SDSS galaxies from \cite{SDSS_2011ApJS..193...29A} as the binned background while the right panel background shows \HII regions from the CALIFA survey \citep{EspinosaPonce_2020MNRAS.494.1622E}. Our regions show a deficiency between 0.5-1.0 dex from the expected value of S2 for a set value of the ${\rm{[OIII]/H}}\beta$ ratios, regardless of which sample we compare to. Our observations are consistent with those observed by \cite{Wang_1997ApJ...491..114W} albeit for galaxies. They inferred that this [SII] deficiency is caused by the leaky \HII regions where Lyman-continuum (LyC) photons are escaping into the surrounded ISM. Since [SII] is mostly produced at the boundary between the ionized and neutral zones of \HII regions, the width of our slits, $1^
{\prime\prime}$ or 0.378 kpc, may not encompass the entirety of the [SII] produced due to the interaction of LyC photons with neutral gas \citep{Pellegrini_2012ApJ...755...40P}. For this reason, we refrain from using any indicator involving [SII].

\subsection{Spectroscopic Observations of Neighboring Dwarf Galaxy} \label{sec:companion}
\begin{figure}[h!]
\plotone{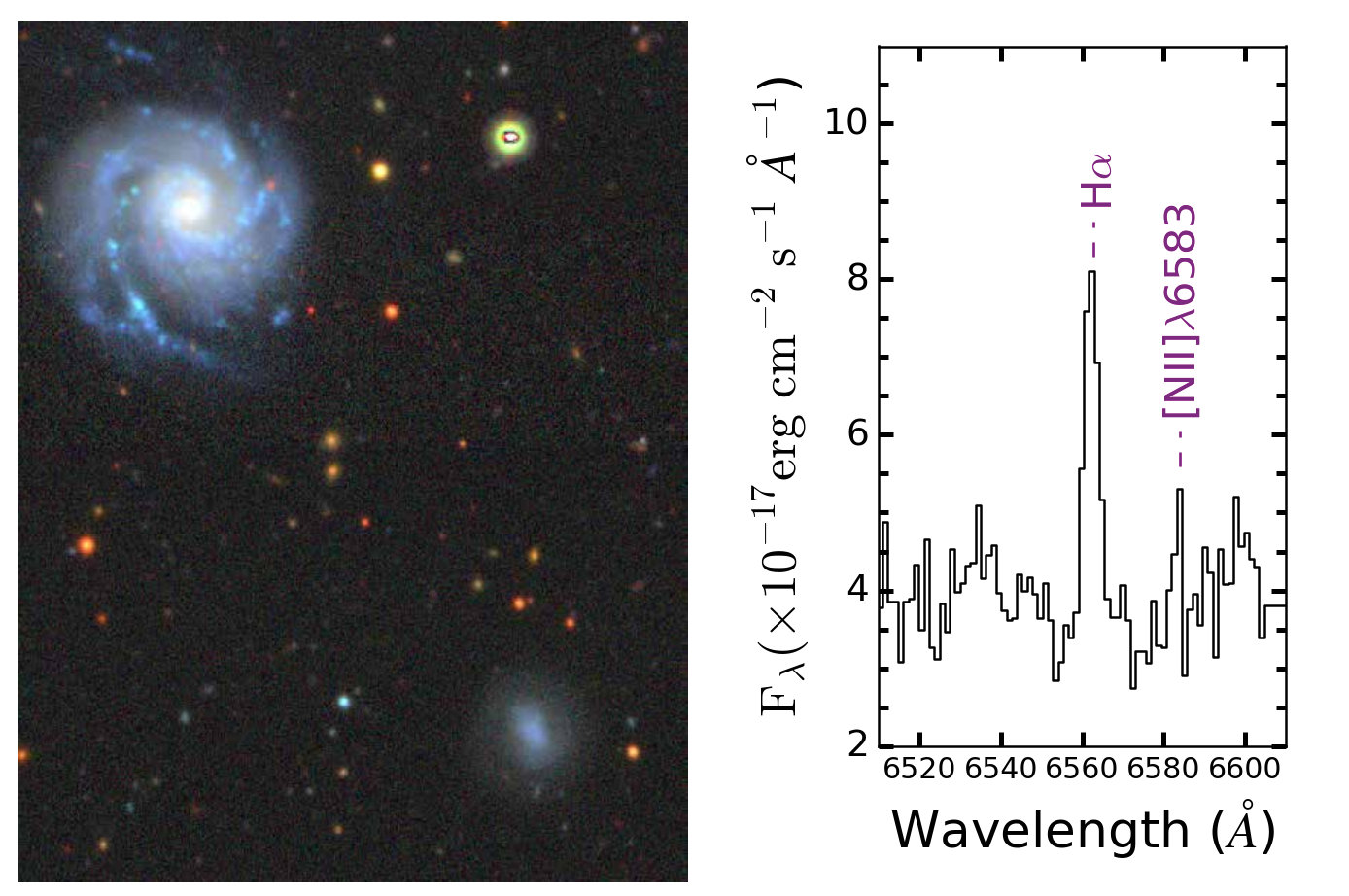}
\caption{(left) Optical image of NGC 99 (top left) and its companion J002353.17+154356.8 (bottom right) from the DESI Legacy Survey \citep{Dey_2019AJ....157..168D}. The projected distance between the objects is 57 kpc. (right) The optical spectrum showing the H$\alpha$ and [\ion{N}{2}]$\lambda$6583 emission lines of J002353.17+154356.8 obtained with the MMT using the Binospec instrument. The redshift was measured to be 0.01747.}
\label{fig:companion}
\end{figure}

SDSS J002353.17+154356.8 is a dwarf galaxy located 57 kpc away from \object{NGC 99} in projection. The left panel of Figure \ref{fig:companion} shows an optical image of \object{NGC 99} with SDSS J002353.17+154356.8 near the bottom right corner of the image. We detail its properties in Table \ref{tab:companion properties}. While observing \object{NGC 99} with the MMT as described in Section \ref{subsec:sample}, we also placed a slit on the center of SDSS J002353.17+154356.8 oriented along its major axis and obtained a spectrum with measurable emissions lines present. The right panel of Figure \ref{fig:companion} shows a cutout of the optical spectrum of the dwarf galaxy. The spectrum was reduced using the same procedure as the spectra in \object{NGC 99}. By measuring the center of the H$\alpha$ emission line, we found a redshift of 0.01747, which is slightly lower than the redshift of \object{NGC 99}. The close proximity of the dwarf galaxy leads us to believe it is a satelli
te of \object{NGC 99}. Shifting the spectrum into \object{NGC 99}'s frame of reference, we measured a velocity offset of -74.51~km s$^{-1}$ with respect to \object{NGC 99}. We calculated that the dwarf galaxy has an N2 value of -0.46 corresponding to 12+log(O/H) = 8.71  \citep{Curti_FMR_2020MNRAS.491..944C}.  

We utilize the $\rm M_{\star}$ map from \cite{Padave_2024ApJ...960...24P} and the \HI 21\,cm image from Gim el al. (in-prep) to measure the $\rm M_{\star}$ and \HI mass of the dwarf galaxy to be $8.62\times10^8~\rm M_\odot$ and $5.8\times10^9~\rm M_\odot$, respectively. The stellar mass of J002353.17+154356.8 is comparable to the Small Magellanic Cloud (SMC; $\rm M_{\star, SMC} = 3.1\times10^8~ M_\odot$) but has a \HI mass an order of magnitude higher than the SMC's gas mass ($\rm M_{gas, SMC} = 4.2\times10^8~ M_\odot$; \citealt{Besla_2015arXiv151103346B}). J002353.17+154356.8 does not show any signs of tidal disruption due to its host galaxy, so we can conclude that the presence of this dwarf galaxy does not affect any of the properties of \object{NGC 99}.

\section{Discussion}\label{sec:discussion}

\subsection{Metallicity Radial Gradient Comparison}
\begin{figure*}[]
\begin{minipage}{0.49\textwidth}
     \centering
     \includegraphics[width=.97\linewidth]{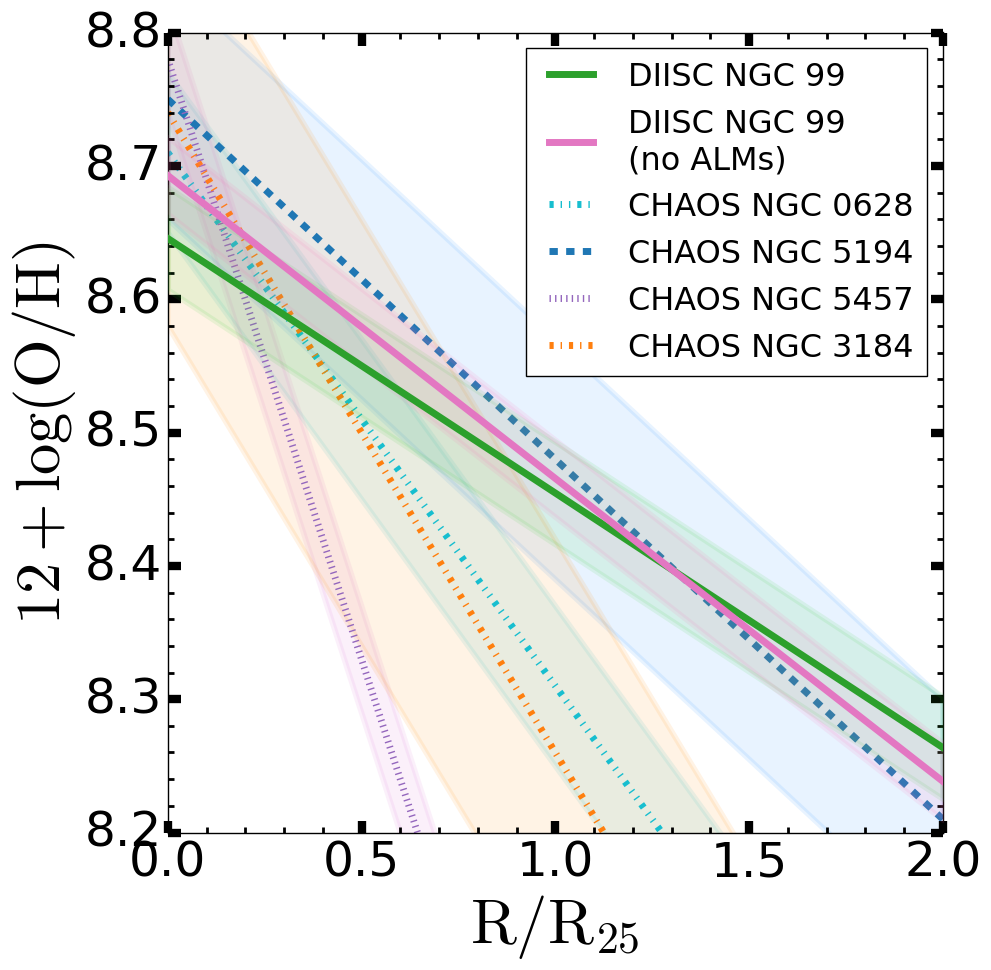}
\end{minipage}
\begin{minipage}{0.49\textwidth}
     \centering
     \includegraphics[width=.89\linewidth]{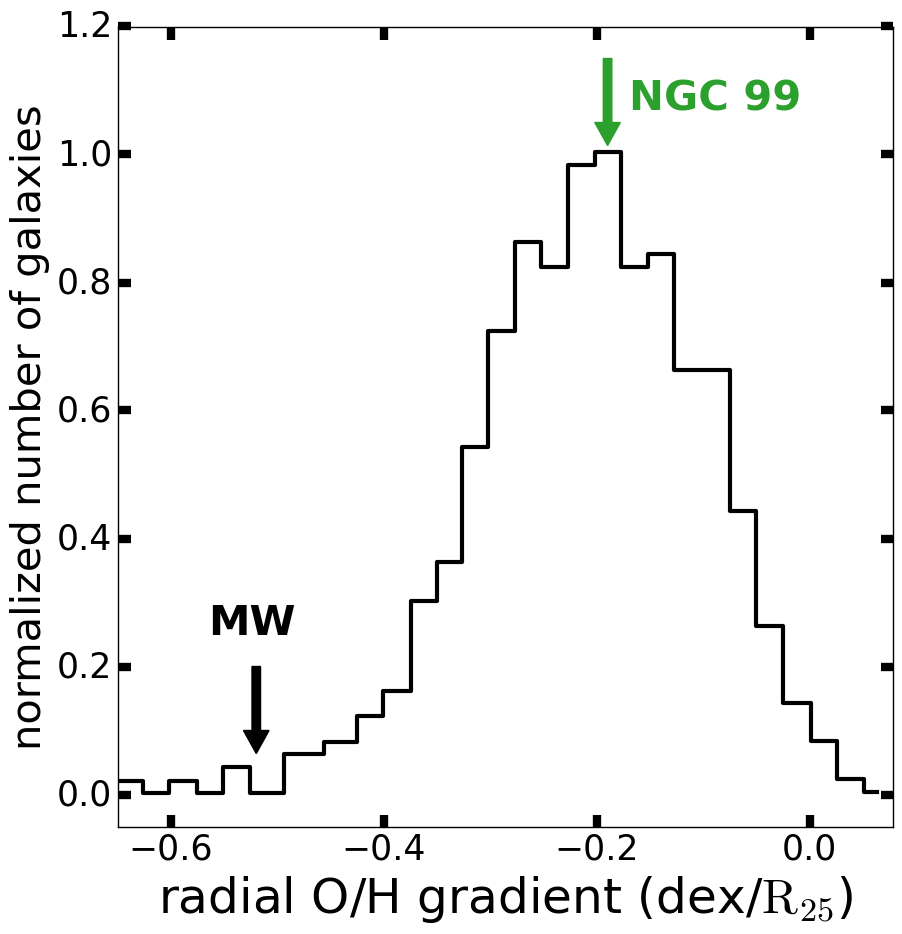}
\end{minipage}
\caption{(\textit{Left}) The metallicity radial metallicity profiles found from our 26 \HII regions in dex/$R_{25}$ in comparison to the four CHAOS galaxies from \cite{Berg_2020ApJ...893...96B}. NGC 99 (green and pink solid lines) has a shallower slope than all of the CHAOS galaxies. The shaded regions indicate the $1\sigma$ error. (\textit{Right}) A histogram of metallicity gradients in 504 galaxies found by \cite{Pilyugin_2023A&A...676A..57P}. The green arrow indicates where NGC 99 fits in this distribution. The black arrow indicates the metallicity gradient observed in the Milky Way.}
\label{fig:radial_gradient_comparison}
\end{figure*}

We compared our radial metallicity gradient to four galaxies from the CHemical Abundances Of Spirals (CHAOS) project that are similar in mass to NGC 99. \cite{Berg_2020ApJ...893...96B} measured the radial metallicity gradient of NGC 0628, NGC 3184, NGC 5194, and NGC 5457 using the direct-method to measure the metallicity of multiple \HII regions observed with the Multi-Object Double Spectrographs (MODS; \citealt{Pogge_2010SPIE.7735E..0AP}) on the Large Binocular Telescope (LBT). The left panel of Figure \ref{fig:radial_gradient_comparison} shows the metallicity distribution of \object{NGC 99} and the four CHAOS galaxies as a function of galactocentric radius normalized by each galaxy's isophotal radius (R$_{25}$). In comparison to \object{NGC 99}, the four CHAOS galaxies are less massive ($10.0 < \rm log~M_\star(M_\odot) < 10.5$) and much closer ($7.2 < \rm D(Mpc) < 11.7$). \object{NGC 99} has a shallower slope than every CHAOS galaxy, even when masking the ALM \HII regions (
pink solid line). It should be noted that our observations of \HII regions in \object{NGC 99} only probe galactocentric radii between 2.3 kpc $\rm < R <$ 18.6 kpc or $0.2 < \rm R/R_{25} < 1.6$, in terms of R$_{25}$. Except for NGC 5457, observations of \HII regions in the CHAOS galaxies also probe galactocentric radii down to similar radii ($\sim$0.1 $\rm R/R_{25}$). Therefore, our linear fit and the fits from 3 of the CHAOS galaxies may be unreliable at R$<$0.2 $\rm R/R_{25}$ as the fit may deviate from linear at lower radii. Additionally, the presence of the two ALM \HII regions lowers the steepness of our slope. As noted in Section \ref{sec:metallicity}, removing regions 15 and 19 from the fit increased the steepness of the metallicity gradient.

\indent Other studies with a large sample size have established a characteristic oxygen abundance gradient in galaxies \citep{Vila-Costas_1992MNRAS.259..121V,Sanchez_2014_slope_A&A...563A..49S, Belfiore_2017MNRAS.469..151B}. Recently, \cite{Pilyugin_2023A&A...676A..57P} measured the radial metallicity gradient of 504 galaxies, of which 451 are MaNGA galaxies and 53 are nearby galaxies. The right panel of Figure \ref{fig:radial_gradient_comparison} shows a histogram of the slopes measured by \cite{Pilyugin_2023A&A...676A..57P} with the black arrow showing where the Milky Way falls in the distribution. The mean of this distribution is -0.20 dex/$\rm R_{25}$ with a scatter of -0.10 dex/$\rm R_{25}$. The green arrow indicates the metallicity gradient of \object{NGC 99} obtained from this work. Our galaxy's gradient (-0.19 dex/$ \rm R_{25}$) approximately matches the mean found by \cite{Pilyugin_2023A&A...676A..57P}. As a part of the CALIFA survey, \cite{Sanchez_2014_slope_A&A...563A..49S} measured the characteristic oxygen abundance of 306 galaxies using a catalog of over 7,000 \HII regions and found a characteristic gradient of -0.16 dex/$ \rm R_{25}$ with a dispersion of 0.12 dex/$ \rm R_{25}$. The gradient of \object{NGC 99} falls within 1$\sigma$ of the characteristic slope found by \cite{Sanchez_2014_slope_A&A...563A..49S}. \cite{Belfiore_2017MNRAS.469..151B} and \cite{Boardman_2021MNRAS.501..948B} showed that the metallicity gradient of a galaxy depends on its stellar mass and size. For a galaxy with a total stellar mass of $\rm log_{10}(M_\star / M_\odot) = 10.6$, \cite{Belfiore_2017MNRAS.469..151B} found that the corresponding metallicity gradient will be about $-0.14~{\rm dex}/R_e$, which is identical to \object{NGC 99}'s metallicity gradient in terms of its effective radius, $R_e$. Despite the difference in observational methods between multi-slit and integral field units (IFUs), which \cite{Sanchez_2014_slope_A&A...563A..49S} and \cite{Pilyugin_2023A&A...676A..57P} utilize, \object{NGC 99} falls within the dispersion of the distribution of radial gradients found by the IFU studies. 

Although \cite{Berg_2020ApJ...893...96B}, \cite{Pilyugin_2023A&A...676A..57P}, and \cite{Sanchez_2014_slope_A&A...563A..49S} study different samples of galaxies and use different methods of measuring metallicities, it is clear that \object{NGC 99} is comparable to most other star-forming galaxies in the local universe. Therefore, the lessons learned from the study of this singular galaxy should be applicable to most other galaxies in the nearby universe.

\subsection{Chemical Evolution Modeling}
Further indirect evidence of gas accretion can be found by modeling the chemical evolution of the \HII regions. The simplest model is the closed-box model, in which it is assumed that no gas leaves or enters the system, and all the gas present is turned into stars over time. As star-formation occurs, a portion of the gas mass is ``locked'' into lower mass stars and stellar remnants as these objects exist for much longer than the current age of the observable universe. Supernovae explosions return another portion of gas mass to the ISM, increasing the local metallicity. This enrichment process is assumed to occur instantaneously. The yield (y) is used to characterize the amount of enrichment that occurred as it is defined to be the mass in metals returned to the ISM relative to the mass in stars formed for a whole population of stars \citep{Tinsley_1980FCPh....5..287T,Maeder_1992A&A...264..105M}. Since we assume that the initial mass function (IMF) remains constant in time and
 no material enters or leaves the system, the yield produced by star-formation remains constant for the galaxy. 

\begin{figure*}
\begin{minipage}{0.49\textwidth}
    \centering
    \includegraphics[width=.9\linewidth]{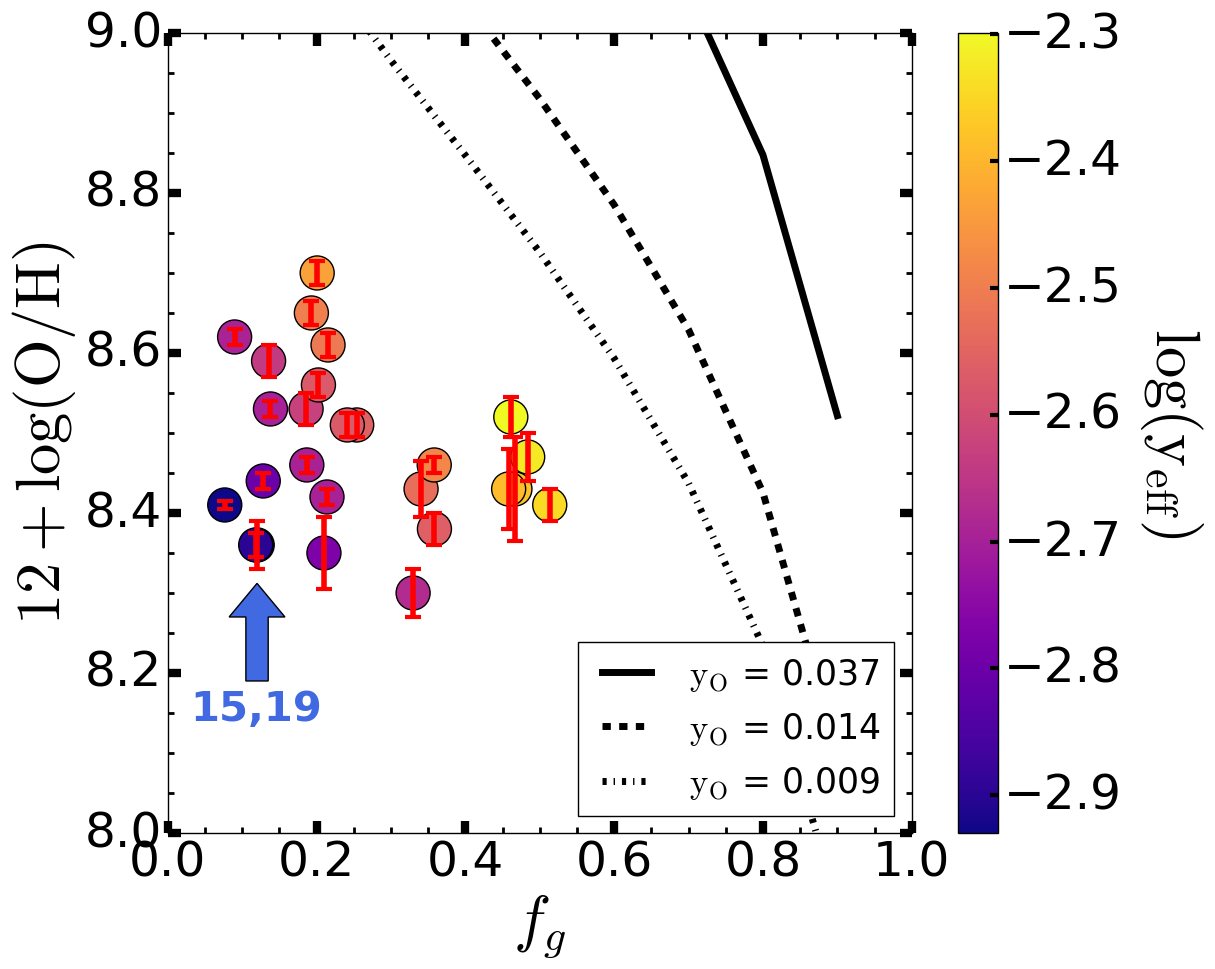}
\end{minipage}%
\begin{minipage}{0.49\textwidth}
    \centering
    \includegraphics[width=.98\linewidth]{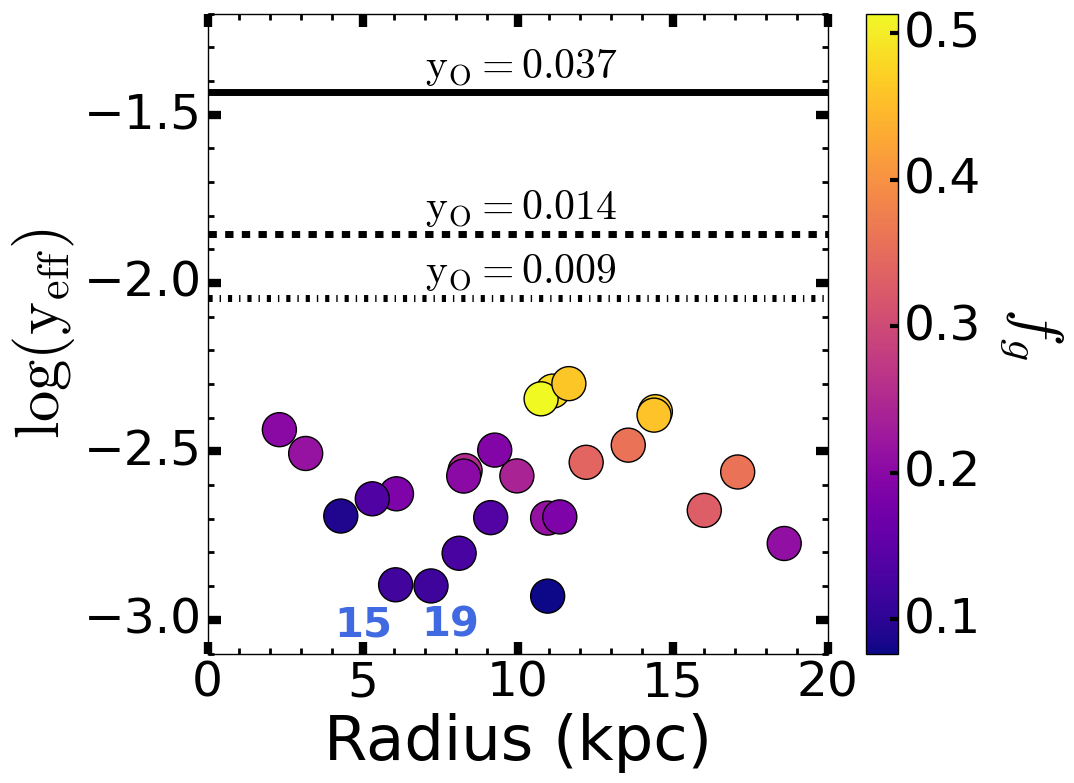}
\end{minipage}
\caption{{\em (Left) } The gas-phase metallicity of the 26 \HII regions is shown as a function of gas fraction with the points color-coded by effective yield, y$\rm _{eff}$. A closed-box model is assumed to calculate effective yield. The lines show the true yield, $\rm y_O$, found by \cite{Leitherer_2014ApJS..212...14L} (dash), and \cite{Vincenzo_Yields_2016MNRAS.455.4183V} (solid and dash dot, respectively). {\em (Right) } For the same \HII regions, the effective yield is now shown as a function of galactocentric radius color-coded by gas fraction. All of the \HII regions have an effective yield lower than the true yield found by \cite{Vincenzo_Yields_2016MNRAS.455.4183V}. The effective yield is constant across the disk of the galaxy. The two ALM \HII regions, numbers 15 and 19, are marked.}
\label{fig:gasfrac}
\end{figure*}

Because the model is simple and makes many assumptions, it serves as a good basis to begin with. We will use it to observe where the assumptions break down and identify possible processes that may be at play. For all models, we assume the galaxy is not pre-enriched with metals. In the closed-box model, the effective yield (y$\rm _{eff}$) is given by the equation

\begin{equation}\label{equ:y_eff}
    {\rm{y_{eff}}} \equiv \frac{Z_{gas}}{{\rm{ln}}(1/f_{gas})}
\end{equation}

\noindent where the gas fraction $f_{gas} \equiv {\rm M_{gas}/(M_{gas} + M_{\star}})$ and $\rm M_{gas}$ and $\rm M_{\star}$ are gas mass and stellar mass, respectively. We utilize the $\rm M_{\star}$ map of NGC 99 created using g- and r-band maps following the method described in \cite{Padave_2024ApJ...960...24P} and estimate the $\rm M_{\star}$ of each \HII region by integrating over the area of the slit. We assume the present-day solar abundances of \cite{Asplund_2021A&A...653A.141A} where hydrogen and oxygen make up $74.38\%$ and $0.58\%$ of the total amount of $\rm M_{gas}$, respectively. Therefore, $\rm M_{gas} = 1.345M_{H}$ following these abundances. We estimate $\rm M_{HI}$ from our VLA \HI 21\,cm surface density image by taking the pixel value at each slit location and multiplying by the area of the slit. The $\rm M_{\star}$ and $\rm M_{HI}$ of each \HII region is shown in Table \ref{tab:FMR}. We use single-dish CO measurements by \cite{Saintonge_2017ApJS..233...22S}
 to estimate a total $\rm M_{H_2}$ of $\rm 10^{9.18}M_\odot$ (after being distance corrected to 79.4 Mpc). The measurements were taken with the Institut de Radioastronomie Millimetrique (IRAM)-30\,m telescope with a half-power beam width (HPBW) of 22$\arcsec$. These measurements are very close to those by \cite{Jiang_2015ApJ...799...92J} who measured a value of $\rm 10^{9.13}M_\odot$ using the Submillimeter Telescope with a HPBW of 33$\arcsec$. This implies that most of the CO is located within 11$\arcsec$ of the center of NGC 99. In the absence of a molecular gas map, we assume the molecular gas to be spread uniformly within a circular region of 11$\arcsec$ ($\approx$4.158 kpc) in radius, which yields a surface density of 27.9 $\rm M_\odot pc^{-2}$. Multiplying the surface density by the area of the slit, the molecular mass, $\rm M_{H_2}$, of each region is estimated to be $\rm 10^{6.99}M_\odot$. Therefore, $\rm M_{H} = M_{HI}+M_{H_2}$ for regions with $R < 4.158$ kpc and $\rm M_{H}
  = M_{HI}$ for regions with $R \geq 4.158$ kpc. Finally, we convert our gas-phase metallicity measured via the N2 index to $\rm Z_{gas}$, the metallicity by mass, via the following equation

\begin{equation}
    \rm log(Z_{gas}) = log(O/H) + 1.07
    \label{equ:GP_to_massFrac}
\end{equation}

\noindent where we assume solar gas-phase metallicity is 8.69 and $\rm Z_{gas} \equiv M_O / M_{gas}$. 

In the closed-box model, the effective yield ($\rm y_{eff}$) should match the ``true" nucleosynthetic yield of oxygen, $\rm y_{O}$, of the system since all the gas present is converted into stars growing the metal content and decreasing the $\rm M_{\star}$. $\rm y_{O}$ can range from 0.009 to 0.037 and is dependent upon metallicity and the assumed IMF \citep{Vincenzo_Yields_2016MNRAS.455.4183V, Barrera-Ballesteros_2018ApJ...852...74B, Weinberg_2023arXiv230905719W}. For this study, we adopt three true yields $\rm y_{O} = 0.009$, $\rm y_{O} = 0.014$, and $\rm y_{O} = 0.037$.

The left panel of Figure \ref{fig:gasfrac} shows a weak inverse correlation between metallicity and gas fraction. Previous studies have found this correlation to be stronger \citep{Barrera-Ballesteros_2018ApJ...852...74B,Lagos_2018MNRAS.477..392L,Tortora_2022A&A...657A..19T}. In particular, \cite{Barrera-Ballesteros_2018ApJ...852...74B} found a tight relation between $f_g$ and 12+log(O/H) using the MaNGA Survey and claim gas fraction plays a vital role in local chemical enrichment. The right panel of Figure \ref{fig:gasfrac} shows that all of the regions have an effective yield lower than that of the true yields (the three lines in the figure). \cite{Edmunds_1990MNRAS.246..678E} showed that $\rm y_{eff}$ should be below the true yield if metals are lost by outflows or the ISM is diluted by inflows of metal-poor gas. Considering the ALM \HII regions previously discussed in Section \ref{sec:metallicity}, the scenario of the inflow of low metallicity gas diluting star-forming re
gions in the galaxy is strengthened by finding $\rm y_{eff} < y_{O}$. The ALM \HII regions (numbers 15 and 19) show particularly low effective yields of -2.90. Additionally, we find that the effective yields of the regions are roughly constant across the galactic disk, with the average being -2.60. 

We now consider the effects of inflow on the chemical evolution of NGC 99 and adopt the accreting box model from \cite{Bovy_prep}. In this model, the effective yield is given by the equation

\begin{equation}
    Z_{gas} = {\rm y_{eff}}\left[1-\exp{\left(1-\frac{1}{f_{gas}}\right)}\right]
\end{equation}

\noindent where it is assumed that the accretion of gas is such that the total amount of gas in the box is constant in time. We tested how using this model would affect the effective yields of the \HII regions and found that $\rm y_{eff}<y_O$, which suggests that accretion cannot be the only mechanism occurring. 

\begin{figure*}
    \centering
    \includegraphics[width = \linewidth]{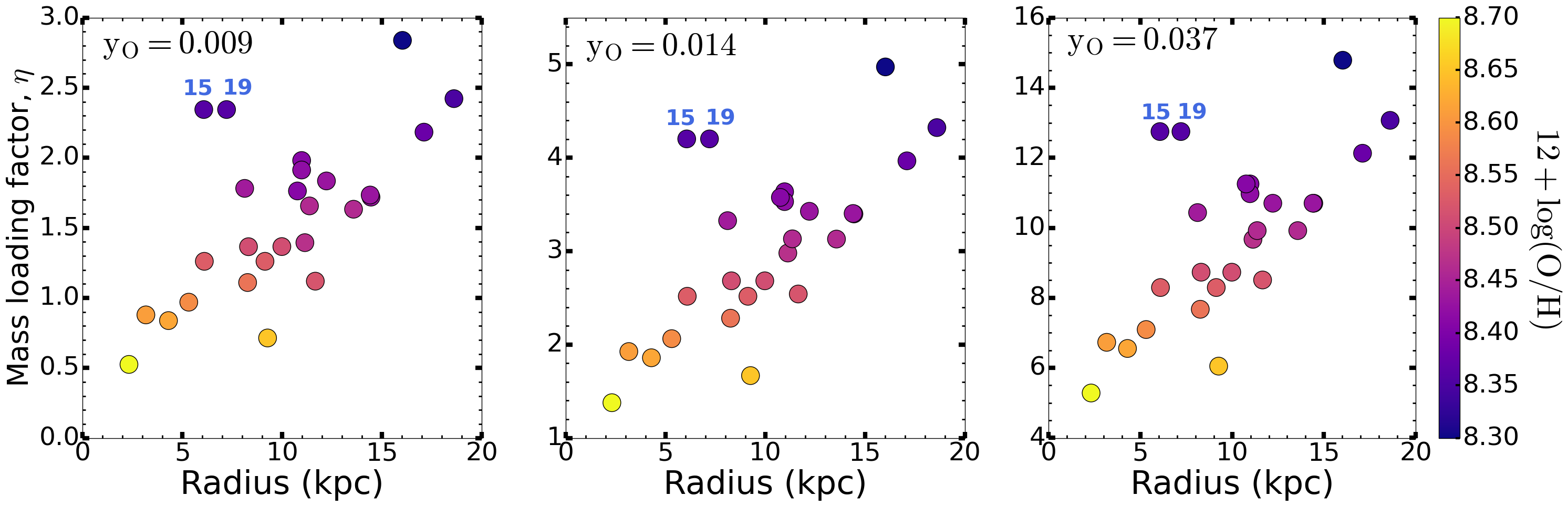}
    \caption{The mass-loading factor $\eta$, modeled using Equation \ref{equ:bathtub}, of 26 \HII regions as a function of galactocentric radius assuming $\rm y_O = 0.009$ (left), $\rm y_O = 0.014$ (middle), and $\rm y_O = 0.037$ (right). The points are color-coded by their gas-phase metallicity. The mass loading factor, as represented in the chemical analysis, gets impacted by metal dilution from low-metallicity gas accretion if $\rm \dot{M}_{inflow} = \dot{M}_{\star}$ and thus deviates from the traditional definition.}
\label{fig:bathtub_model}
\end{figure*}

Chemical evolution models that include both inflows and outflows may offer a more realistic model to describe galaxies like \object{NGC 99}. \cite{Bovy_prep} provides an adjusted form of the accreting box known as the `gas regulatory' or `bathtub' model, which includes outflows  \citep{Tinsley_1973ApJ...186...35T,Tinsley_1978ApJ...221..554T,Bouche_2010ApJ...718.1001B, Lilly_2013ApJ...772..119L}. In this model, it is assumed that the star-formation rate is equal to the inflow rate ($\rm \dot{M}_{inflow}$) of material or $\rm \dot{M}_{\star} = \dot{M}_{inflow}$, which has been found to be broadly correct on global scales based on simulations and modeling \citep{Fin_Dave_2008MNRAS.385.2181F, Bouche_2010ApJ...718.1001B, Dayal_2013MNRAS.430.2891D, Peng_2014MNRAS.443.3643P}. The model can be described by the equation

\begin{equation}\label{equ:bathtub}
    Z_{gas} = \frac{{\rm y}}{1+\eta}\biggl\{1- \exp{\biggl([1+\eta]\left[1-\frac{1}{f_{gas}}\right]\biggr)}\biggl\}
\end{equation}

\noindent where the mass loading factor, $\eta$, is defined as $\eta = \rm \dot{M}_{outflow}/\dot{M}_{\star}$. Setting $\rm y = y_{O}$, we solve Equation \ref{equ:bathtub} for $\eta$ to see how the mass-loading factor varies as a function of galactocentric radius. We find that most of the \HII regions show a higher mass-loading factor at a higher galactocentric radius, except for two of them, as seen in Figure \ref{fig:bathtub_model}. For their given radius, the two ALM \HII regions (regions 15 and 19) both show a high $\eta$ for their respective position in the galaxy. 

Figure \ref{fig:bathtub_model} shows that $\eta$ can vary from 0.5-15 depending on the value of $\rm y_O$. Mass-loading factors of this scale are typically found in starbursting galaxies \citep{Heckman_2015ApJ...809..147H}. Following \cite{Rodighiero_2011ApJ...739L..40R}, a galaxy is defined to be starbursting if SFR$\rm _{global}$/SFR$\rm _{MS}$ $>$ 4, where SFR$\rm _{MS}$ is the global SFR from the star-forming galaxy main sequence (MS). We utilize the time-dependent MS equation of \cite{Speagle_2014ApJS..214...15S} to determine SFR$\rm _{MS}$ = 1.64 $\rm M_\odot~yr^{-1}$. Therefore, SFR$\rm _{global}$/SFR$\rm _{MS}$ = 1.74, so \object{NGC 99} is not currently in a starbursting phase. For cases where $\eta$ is much greater than 1, the mass-loading factor becomes nonphysical for a galaxy of this mass \citep{Chisholm_2017MNRAS.469.4831C,Xu_2022ApJ...933..222X}. Therefore, the most likely value of true yield is around $\rm y_O = 0.009$. Regardless of $\rm y_O$, a correlation b
etween $\eta$ and R is evident. $\eta$ increases with radius and shows an inverse correlation with stellar density. The mass-loading factor is highest in the outermost \HII regions, which is contradictory to the traditional expectation for $\eta$. Instead, the high mass-loading factors are an artifact of the model we assumed. Equation \ref{equ:bathtub} shows that $\eta$ and $\rm Z_{gas}$ are roughly inversely related; therefore, a lower metallicity can result in higher mass-loading factor values. The color coding in Figure \ref{fig:bathtub_model} confirms that $\eta$ has an inverse relation with gas-phase metallicity. The outer \HII regions with lower metallicities have the highest mass-loading factors. This suggests that $\eta$, in this situation, is tracing gas dilution due to inflow.

Both processes, outflow of metal-rich gas and inflow of low metallicity gas accretion, can lower the metallicity of the \HII regions, causing their mass-loading factor to be high. Even if NGC~99 had an outflow (which we cannot be certain about), it would have $\eta$ proportional to the star-formation rate, which can be approximated by the stellar mass surface density today if a significant fraction of the stellar mass was created in that burst. Since that is not the case, the most likely process leading to the increasing $\eta$ trend with radius would be the accretion of lower metallicity gas. This is supported by the fact that gas accretion is believed to occur mostly at the outskirts of the galactic disks along their major axis \citep{Peroux2020a}.

We acknowledge that the bathtub chemical evolution models are not sophisticated enough to precisely model the processes of mixing and diffusion. 
Another limitation of the model is that it does not account for differential accretion, i.e., when $\rm \dot{M}_\star \neq \dot{M}_{inflow}$ and $\rm \dot{M}_{inflow}$ is independent of $\rm \dot{M}_{\star}$. Nevertheless, this model helps us understand the balance of gas outflow and inflow in a qualitative manner, which is what we advise the reader to take away from this discussion.

\subsection{Estimation of Accreted Mass}
We can estimate the accreted amount of low metallicity gas needed to dilute the original gas present in the two ALM regions to the metallicity value we observe today. The accreted mass $\rm M_{H,acc}$ can be given by the following equation

\begin{equation}
    \label{equ:mass_acc}
    \rm M_{H,acc} = M_{H,now} - M_{H,orignal}
\end{equation}

\noindent where $\rm M_{H,now}$ is the hydrogen mass of the region now and $\rm M_{H,original}$ is the original hydrogen mass of the system. We assume that the metals in the \HII regions are primarily made up of oxygen so, the mass in metals we currently see $\rm M_{O,now}$ can then be given by  

\begin{equation}
    \label{equ:oxy_mass_acc}
    \rm M_{O,now} = M_{O,original} + M_{O, acc}
\end{equation}

\noindent where $\rm M_{O,original}$ is the original oxygen mass of the region before dilution and $\rm M_{O,acc}$ is the oxygen mass that was accreted in. If we assume that the gas accreted was pristine, that is $\rm M_{O, acc}=0$, then $\rm M_{O,now} = M_{O,original}$. The oxygen mass can then be estimated using metallicity and \HI 21\,cm mass values using $\rm M_{O, now} = M_{H,now}\times Z_{now, ALM}$ where $\rm Z_{now, ALM}$ is the measured metallicity of the two ALM regions. We convert 12 + log(O/H) to $\rm Z_{now, ALM}$ using Equation \ref{equ:GP_to_massFrac}. We estimate the mass of oxygen originally present in regions 15 and 19 to be 8.8$\times10^3~\rm M_\odot$ and 8.2$\times10^3~ \rm M_\odot$, respectively. The original amount of hydrogen is $\rm M_{H,original} = M_{O,original}/Z_{original}$ where $\rm Z_{original}$ is the metallicity of the \HII regions before dilution. $\rm Z_{original}$ is estimated using the linear fitted metallicity gradient from the N2 indicat
or where we excluded the ALM regions from the fit, which can be found in Table \ref{tab:linearfits}. $\rm Z_{original}$ for the two regions is 8.57 and 8.55. Thus, $\rm M_{H,original}$ of the \HII regions is 2.01$\times10^6~\rm M_\odot$ and 1.98$\times10^6~ \rm M_\odot$.  Using Equation \ref{equ:mass_acc}, we estimate the accreted hydrogen mass needed to dilute the systems to be 1.26$\times10^6~\rm M_\odot$ and 1.08$\times10^6~ \rm M_\odot$ for regions 15 and 19, respectively.

These accreted masses should be considered lower limits since we assume that the accreted gas is pristine. The masses of the accreted gas clouds are similar to those found in high-velocity clouds around the Milky Way \citep{Putnam_2012ARA&A..50..491P}.

\subsection{CGM of NGC~99}

\begin{figure*}[]
    \centering
    \includegraphics[width = 6.5in]{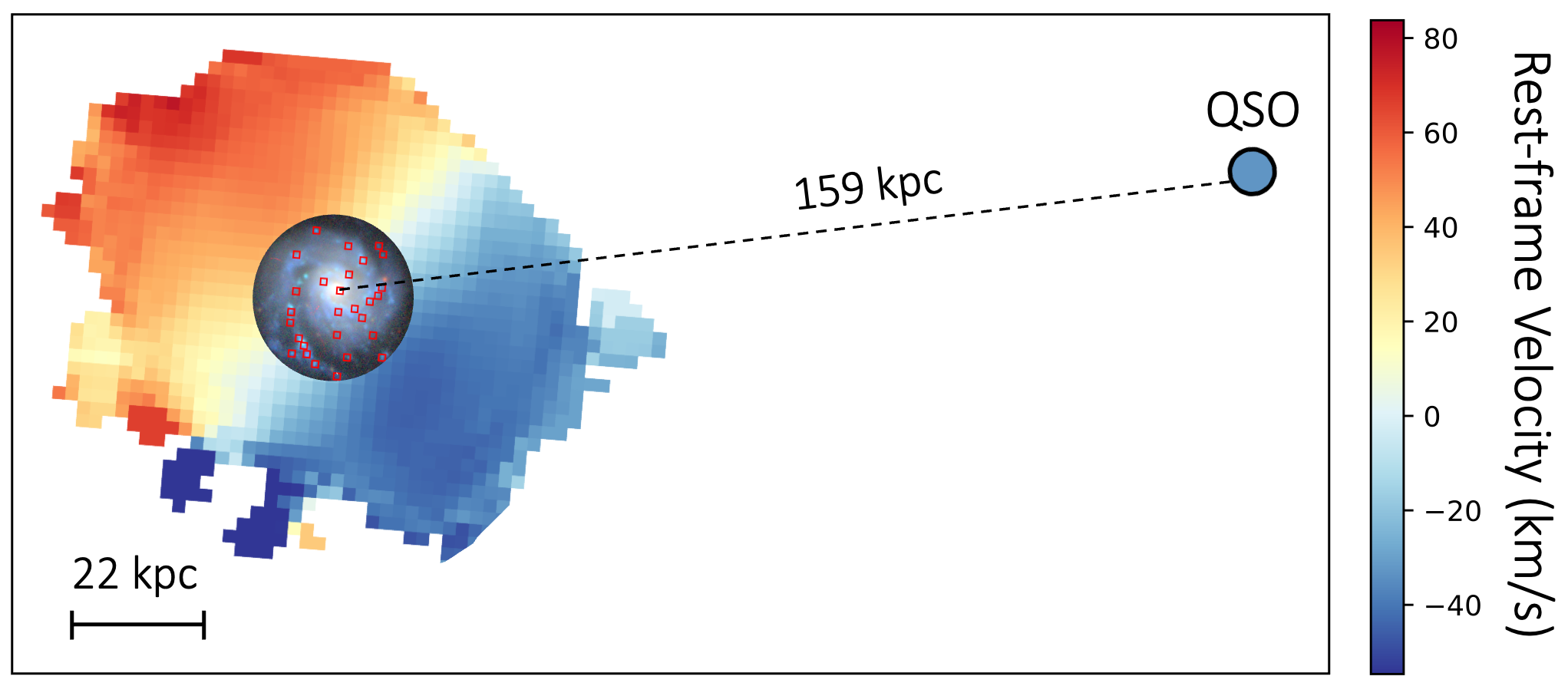}
    \caption{An optical image of NGC 99 overlaid with VLA \HI 21\,cm map showing extended \HI emission beyond the optical disk of the galaxy. The spatial location of QSO (J0023+1547) is shown relative to NGC 99, with the color of the point corresponding to the velocity centroid of the dominant component of the Ly$\alpha$ line. The primary component of the CGM is co-rotating at roughly the same velocity as the ISM closest to the sightline. This suggests a large disk extending out to the CGM.}
    \label{fig:CGM Co-rotation}
\end{figure*}

The absorption spectra at the position of the QSO J0023+1547 trace four clouds in Ly$\alpha$, the strongest of which matches the velocity of the \HI 21cm emitting ISM within the disk of the NGC~99. Figure~\ref{fig:CGM Co-rotation} shows the ISM kinematics as traced by \HI 21cm with the galaxy's optical map overlaid. The position of the QSO is indicated with a filled circle that is color-coded to the velocity of the strongest component (A). The strongest component (see Table~\ref{tbl:CGM}) is corotating with the ISM with an uncertainty of less than 5 km s$^{-1}$. The large synthesized beam  ($50.5^{\prime\prime} \times 47.5^{\prime\prime}$) of the VLA smears the small-scale velocity structure of the \HI data. Therefore, the precise measurement of velocity at the edge of the \HI disk is not possible with our data. 

The corotating component has a column density, $\rm N(HI) \ge 10^{14.4}~cm^{-2}$, which is more than an order of magnitude higher than all the other three components put together, indicating that most of the neutral material is corotating. The three weaker components are non-corotating, i.e., their velocities are reversed from what is seen on the blueshifted side of the galaxy. If the covering fraction of the strong component is large, then we are looking at the extended disk material associated with the disk. Galaxy simulations have seen the presence of large extended corotating disks \citep{Stewart11}.
Recent simulations by \citet{Hafen22} also found that gas accreting into thin-disk galaxies in the FIRE simulations is dominated by rotating
cooling flows. The hotter accreting gas cools down to $\rm 10^4~K$, and its geometry transitions from a quasi-spherical distribution to a cool extended disk.
 
The CGM gas distributions and kinematics suggest two possible entryways for gas into the disk of the galaxies. First is through the extended disk traced by the dominant component. While the gas in the extended disk is much further from where the ALMs are seen (at galactocentric radii of $<$10~kpc), the gas may flow inwards without much metal mixing for galaxies with low gas dispersion ($\sigma_g \sim \rm 14~km/s$) through radial flows in the ISM \citep{Sharda21}. 
The steep increase in the mass loading fraction as a function of radius (see Figure~\ref{fig:bathtub_model}) 
suggests that gas dilution via accretion is increasingly important at large radii. This scenario is consistent with the radial flow of low-metallicity gas along the disk. 
\cite{Padave_2024ApJ...960...24P} find no correlation between the neutral gas content of the CGM and inside-out disk growth, suggesting that gas likely transverses large distances to get to the inner stellar disk before forming stars.

The second pathway is through the halo vertically to the disk. High-velocity clouds (HVC) are one such example that show an overall infall that is believed to support star formation in the Milky Way \citep{Lehner22}. Our estimate of the inflowing cloud masses needed to produce the ALMs falls within the HVC mass range of the Milky Way  \citep{Putnam_2012ARA&A..50..491P} and anomalous velocity clouds analogous to intermediate-velocity clouds seen in NGC~4321 \citep{Gim21}.

No metal absorbers associated with the hydrogen clouds were detected (Table~\ref{tbl:CGM}). This is not surprising as even at solar metallicity, the metal lines are expected to be very weak for gas clouds of $\rm N(HI) \ge 10^{14-15}~cm^{-2}$. We do not expect the \Lya absorbers to be tracing highly ionized gas as we did not detect high-ionization species such as \ion{Si}{4} and \ion{N}{5}. 

\subsection{Fundamental Metallicity Relation on Resolved Scales}

\cite{Lequeux_1979A&A....80..155L}, \cite{Zaritsky_1994ApJ...420...87Z}, and  \cite{Tremonti_2004ApJ...613..898T} established a positive correlation between global metallicity and the $\rm M_\star$ of galaxies, known as the mass-metallicity relation (MZR).  Modern studies, such as MaNGA, have shown that the MZR evolves over cosmic time (\citealt{MaNGA_2015ApJ...798....7B,Camps_2022ApJ...933...44C}). In addition, observations have shown that metallicity and $\rm M_\star$ depend on SFR \citep{Lara-Lopez_2010A&A...521L..53L,Mannucci_2010MNRAS.408.2115M, Curti_FMR_2020MNRAS.491..944C}. The relation between the three properties is known as the Fundamental Metallicity Relation (FMR). \cite{Lara-Lopez_2010A&A...521L..53L} and \cite{Mannucci_2010MNRAS.408.2115M} found that at a fixed $\rm M_{\star}$, a higher SFR yields a lower metallicity. Additionally, as $\rm M_{\star}$ increases, so do metallicity and SFR. Results from \cite{Rosales-Ortega_2012ApJ...756L..31R}, which were later c
onfirmed by \cite{BarreraBallesteros_2016MNRAS.463.2513B}, established that the MZR and FMR also occur on localized (sub-kpc) scales. We will further explore the FMR here.

We derived the SFR from the ${\rm{H}}\alpha$ emission line using

\begin{equation}
    \label{equ:SFR}
    \rm log ~SFR (M_\odot~year^{-1}) = log~ L_x - log~ C_x
\end{equation}

\noindent where $\rm log~C_x = 41.27$ and $\rm L_x$, in units of $\rm ergs~s^{-1}$, is the ${\rm{H}}\alpha$ luminosity derived from the integrated and dust-corrected ${\rm{H}}\alpha$ flux at a distance of 79.4 Mpc \citep{Hao_2011ApJ...741..124H,Murphy_2011ApJ...737...67M,Kennicutt_2012ARA&A..50..531K}. We convert $\rm M_{\star}$ and SFR to stellar mass surface density ($\rm \Sigma_\star$) and star-formation rate surface density ($\rm \Sigma_{SFR}$), respectively, by dividing $\rm M_{\star}$ and SFR by the area of each slit, 0.35 kpc$^2$.

\begin{figure*}
    \begin{interactive}{animation}{myfigs/Paper_rFMR_N2_v6.mp4}
    \plotone{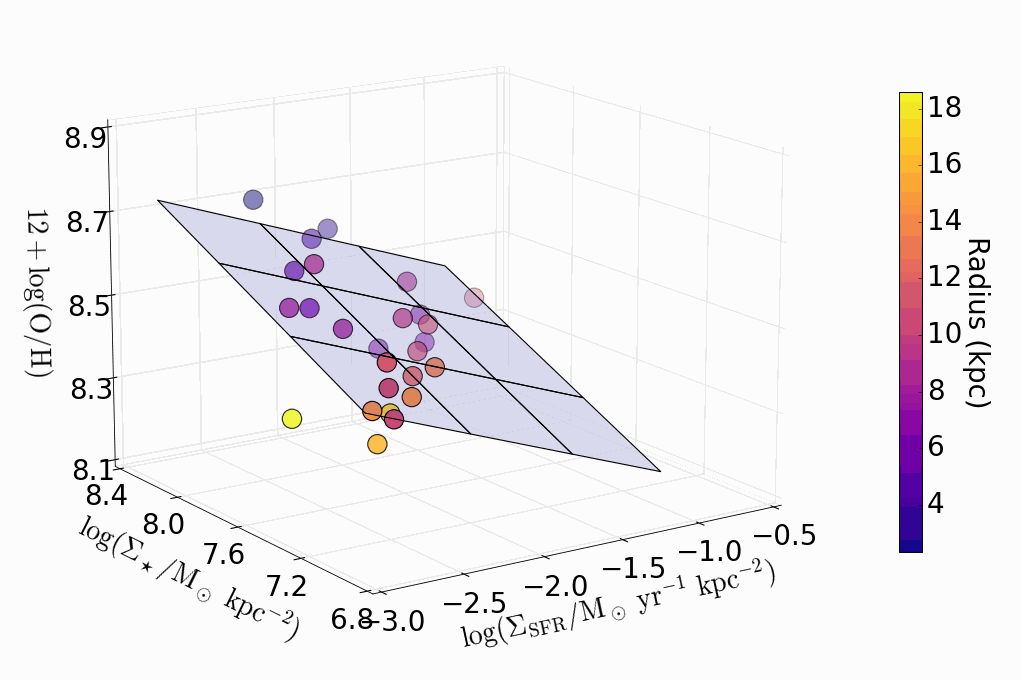}
    \end{interactive}
    \caption{3D visualization of the resolved Fundamental Metallicity Relation for the 26 \HII regions in the galaxy NGC 99 color-coded by galactocentric radius. The x-axis and y-axis are the stellar mass and star-formation rate surface densities, respectively, for each region. Gas-phase metallicities from the N2 indicator are shown on the z-axis. The points are well fit my a linear plane described by Equation \ref{equ:FMR}. An animated version of this figure is available online, where we rotate around the z-axis for one full rotation in an 18-second animation.}
    \label{fig:3D_FMR}
\end{figure*}

We investigate the relationship between $\rm \Sigma_\star$, $\rm \Sigma_{SFR}$, and metallicity by plotting the data in three-dimensional space to gain a better visual representation. Figure \ref{fig:3D_FMR} shows the $\rm \Sigma_\star$, $\rm \Sigma_{SFR}$, and 12+log(O/H) parameter space with the filled circles color-coded based on their galactocentric radius in kpc. We fit a plane to the data and find that the relationship between the three properties is described by the equation:

\begin{equation}\label{equ:FMR}
  \begin{aligned}
    \rm 12 + log(O/H) & =\\
      & (-0.144\pm0.044)\times\rm log(\Sigma_{SFR}) \\
      & + (0.225\pm0.050)\times\rm log(\Sigma_{\star}) \\
      & + (6.443\pm0.455)
  \end{aligned}
\end{equation}

\noindent Figure \ref{fig:3D_FMR} is available as an online video where we rotate about the z-axis one time so that the entirety of the plane is seen. The coefficients in front of the two surface density parameters describe how and to what strength gas-phase metallicity depends on local $\rm \Sigma_\star$ and $\rm \Sigma_{SFR}$. The resolved FMR (rFMR), given by Equation \ref{equ:FMR}, shows that for a given $\rm \Sigma_\star$, metallicity relates inversely with $\rm \Sigma_{SFR}$. Moreover, for a given $\rm \Sigma_{SFR}$, metallicity depends proportionally on $\rm \Sigma_\star$. The value of the coefficients, -0.144 and 0.225, for $\rm \Sigma_{SFR}$ and $\rm \Sigma_\star$, respectively, reveals that the metallicity of the \HII regions has a stronger dependence on $\rm \Sigma_\star$ than $\rm \Sigma_{SFR}$. The scatter in metallicity around the best-fit plane is $\sigma_{FMR} = 0.07$ dex, which is comparable to the scatter of the surface fit by \cite{Curti_FMR_2020MNRAS.491..944C}. \cite{Baker_2023MNRAS.519.1149B} recently found similar results using 56,000 \HII regions from the MaNGA survey and measure $\sigma_{FMR} = 0.06$ dex.  

Our rFMR equation presents the resolved FMR at spatial scales of 4 arcseconds (1.5 kpc), which provides a view of kpc scale physics that may be responsible for the observed relationship. The spatial resolution we achieve is comparable to some of the IFU surveys; however, the multi-object spectrograph allows us to sample a much larger region of the galaxy going well beyond R$_{25}$. Moreover, the size of our Binospec slits encompasses the gas and stars within the observed regions allowing us to probe their chemical characteristics. Studies on the rFMR have been done using spatially resolved data from the MaNGA sample, which samples spatial scales of about 1.5 kpc \citep{BarreraBallesteros_2016MNRAS.463.2513B,SanchezMenguiano_2019ApJ...882....9S, Teklu_2020ApJ...897...61T}. We will increase our sample size and spatial resolution in a future study.

\section{Summary} \label{sec:conclusion}
In this pilot study, we study the properties of 26 \HII regions in the galaxy \object{NGC 99}. We measured the radial metallicity gradient from strong-line metallicity indicators of N2 and O3N2 to investigate indirect evidence of gas accretion. Additionally, we analyze the chemical evolution of the systems by assuming a closed-box, accreting box, and accreting/outflow box models. We measure an effective yield and mass-loading factor for each region. Finally, we derive a resolved Fundamental Metallicity Relation by connecting the metallicity information to $\rm \Sigma_\star$ and $\rm \Sigma_{SFR}$. The results are summarized below.

\begin{enumerate}
    \item Using oxygen gas-phase metallicities calibrated using the N2 metallicity index, we measure the radial metallicity gradient of \object{NGC 99} to be -0.017 dex kpc$^{-1}$ or, in terms of $\rm R_{25}$, -0.19 dex/$\rm R_{25}$. Our radial gradient matches with the peak of the distribution of radial gradients found by other studies \citep{Berg_2020ApJ...893...96B, Pilyugin_2023A&A...676A..57P}.
    \item We discover two anomalously low metallicity regions with a high difference between the \HI and ${\rm H}\alpha$ gas line-of-sight velocities. We believe that these two regions are an indirect sign of accretion of low metallicity gas from the CGM that diluted the gas used to form the stars in these regions.
    \item Assuming the closed-box model, we find that the effective yield of each observed \HII region is below the true nucleosynthetic yield. Additionally, the effective yields are roughly constant with galactocentric radius. These results lead us to believe that the galaxy has a varying degree of inflows and outflows.
    \item When inflow and outflows are incorporated, we find that the \HII regions have a mass-loading factor between 0.5 and 15 depending on the choice of y$\rm _O$. Additionally, we suggest higher accretion rates of low metallicity gas in the outskirts of the galaxy as the reason for higher $\eta$ values at higher radii. This occurs as a consequence of using the bathtub model, which does not account for differential accretion.
    \item We find \Lya absorbers associated with the CGM of the galaxy at an impact parameter 159~kpc. The strongest component of the absorption feature kinematically matches the disk kinematics closest to the sightline. This may hint at the presence of an extended disk, assuming that the covering fraction of the absorbing media is high. We suggest that large \HI disks may be a possible pathway for low-metallicity gas to flow into the ISM of this galaxy.      
    \item We derive Equation \ref{equ:FMR} describing the resolved Fundamental Metallicity Relation relating $\rm Z$, $\Sigma_{\star}$, and $\rm \Sigma_{SFR}$. We find that for a given $\Sigma_{\star}$, $\rm \Sigma_{SFR}$ increases with decreasing metallicity. From this result, we believe that the local physics within galaxies affects the global scale physics. Our equation is useful for estimating metallicities in galaxies where only imaging data might be available, which is especially the case for higher redshift galaxies.  
\end{enumerate}

Future work will expand this analysis to other galaxies in the DIISC sample to look for other ALM \HII regions and grow our sample size for the rFMR to improve the precision of the plane. Multiwavelength studies of local galaxies are critical to obtaining a detailed look into how galaxies obtain their gas and how it is processed for star formation.

\begin{acknowledgments}
\textbf{ACKNOWLEDGMENTS}

We thank the referee for their constructive comments. 

This material is based upon work supported by the National Science Foundation Graduate
Research Fellowship Program under Grant No. 2233001. Any opinions, findings,
and conclusions or recommendations expressed in this material are those of the author(s)
and do not necessarily reflect the views of the National Science Foundation.

A.O., S.B., M.P., B.K., H.G., and C.D. are supported by the NSF grants 2108159 and 2009409. M.P., S.B., and R.J. are supported by NASA ADAP grant 80NSSC21K0643. S.B., H.G., and T.H. are also supported by HST grant HST-GO-14071 administrated by STScI, operated by AURA under contract NAS 5- 26555 from NASA. 

The Arizona State University authors acknowledge the twenty-three Native Nations that have inhabited this land for centuries. Arizona State University's four campuses are located in the Salt River Valley on ancestral territories of Indigenous peoples, including the Akimel O’odham (Pima) and Pee Posh (Maricopa) Indian Communities, whose care and keeping of these lands allows us to be here today.

We thank Jackie Monkiewicz, Ben Weiner, Chris Howk, Mirko Curti, and Patrick Kamieneski for their valuable and useful discussions during the course of this work. We thank the support staff at the MMT Observatory, the Steward Observatory, the Vatican Advanced Technology Telescope, the Very Large Array, the National Radio Astronomy Observatory, and the Space Telescope Science Institute for their help with this project. All HST data presented in this paper were obtained from the Mikulski Archive for Space Telescopes (MAST) at the Space Telescope Science Institute. The specific observations analyzed can be accessed via\dataset[DOI: 10.17909/rf6m-3y73]{https://doi.org/10.17909/rf6m-3y73}.

Observations reported here were obtained at the MMT Observatory, a joint facility of the Smithsonian Institution and the University of Arizona.

This work is also partly based on observations with the VATT: the Alice P. Lennon Telescope and the Thomas J. Bannan Astrophysics Facility.

The National Radio Astronomy Observatory is a facility of the National Science Foundation operated under cooperative agreement by Associated Universities, Inc.

\end{acknowledgments}

%

\vspace{5mm}
\facilities{HST, GALEX, MMT (Binospec), VATT, and VLA}


\software{astropy \citep{2013A&A...558A..33A,2018AJ....156..123A,astropy2022ApJ...935..167A},
          CASA \citep{McMullin_2007ASPC..376..127M},
          \texttt{ipython/ jupyter} \citep{Perez_2007CSE.....9c..21P,Kluyver_2016ppap.book...87K},
          \texttt{matplotlib} \citep{Hunter:2007},
          \texttt{NumPy} \citep{harris2020array}, 
          \texttt{pyFIT3D} \citep{Lacerda_2022NewA...9701895L},
          \texttt{SoFiA-2} \citep{Serra_2015MNRAS.448.1922S,Westmeier_2021ascl.soft09005W},
          \texttt{statmorph} \citep{Rodriguez-Gomez_2019MNRAS.483.4140R},
          \texttt{specutils} \citep{specutils}, and 
          \texttt{Python} from \url{https://www.python.org}}



\bibliography{NGC99-ALM}{}
\bibliographystyle{aasjournal}
\vspace{-70mm}
\appendix

\begin{deluxetable*}{CCCCCCCCCCC}[h!]
\rotate
\tablenum{3}
\label{tab:indicators}
\tablecaption{NGC 99 \HII Region Position, Flux, Balmer $\&$ Metallicity Indicator Values}
\tablewidth{0pt}
\tabletypesize{\footnotesize}
\tablehead{
\colhead{\HII} & \colhead{R.A.} &  \colhead{Decl.} & \colhead{Radius} & \colhead{F(H$\alpha$)} & \colhead{F(H$\alpha$)/F(H$\beta$)} &  \colhead{N2} & \colhead{O3N2} & \colhead{S2} & \colhead{${\rm V}_{{\rm H}\alpha}$} & \colhead{${\rm V}_{{\rm HI}}$}\\
\colhead{Region} & \colhead{(degrees)} & \colhead{(degrees)} & \colhead{(kpc)} & \colhead{($\rm erg ~s^{-1} ~cm^{-2}$)} & \colhead{} & \colhead{} & \colhead{} & \colhead{} & \colhead{(km/s)}& \colhead{(km/s)}
}
\startdata
1     & 5.9966 & 15.7598 & 14.4  & 1.27  & 4.44\pm0.41 & -0.966\pm0.127 & 1.2726\pm0.1349 & -0.699\pm0.051 & -66.33 & -26.41 \\
2     & 5.9907 & 15.7627 & 14.4  & 2.56  & 3.37\pm0.31 & -0.974\pm0.098 & 1.4556\pm0.1050 & -0.651\pm0.067 & -23.78 & -35.86 \\
3     & 5.9952 & 15.7621 & 11.7  & 1.71  & 3.19\pm0.14 & -0.802\pm0.048 & 0.8396\pm0.0580 & -0.611\pm0.043 & -39.04 & -29.07 \\
4     & 6.0006 & 15.7620 & 12.2  & 4.26  & 3.98\pm0.38 & -0.975\pm0.069 & 1.5007\pm0.0815 & -0.658\pm0.043 & -17.45 & -13.18 \\
5     & 6.0026 & 15.7620 & 13.6  & 4.73  & 3.65\pm0.16 & -0.919\pm0.024 & 1.3667\pm0.0352 & -0.668\pm0.020 & -3.44 & -3.20 \\
6     & 6.0009 & 15.7628 & 11.4  & 13.13 & 3.81\pm0.17 & -0.917\pm0.025 & 1.2389\pm0.0373 & -0.740\pm0.025 & -10.99 & -3.98 \\
7     & 6.0014 & 15.7634 & 11.0  & 79.82 & 5.14\pm0.15 & -1.006\pm0.017 & 1.4924\pm0.0265 & -0.913\pm0.016 & -33.62 & -3.98 \\
8     & 5.9934 & 15.7662 & 8.3   & 2.79  & 4.29\pm0.21 & -0.823\pm0.035 & 0.9739\pm0.0482 & -0.537\pm0.029 & -47.36 & -25.09 \\
9     & 5.9976 & 15.7658 & 6.1   & 2.67  & 4.45\pm0.28 & -0.776\pm0.039 & 0.8317\pm0.0530 & -0.658\pm0.456 & -28.33 & -7.96 \\
10    & 6.0032 & 15.7662 & 10.0  & 6.72  & 4.17\pm0.10 & -0.826\pm0.021 & 1.0075\pm0.0250 & -0.737\pm0.019 & 23.09 & 10.26 \\
11    & 6.0033 & 15.7678 & 9.1   & 14.58 & 3.80\pm0.23 & -0.778\pm0.025 & 0.9274\pm0.0429 & -0.705\pm0.024 & 48.38 & 14.89 \\
12    & 5.9945 & 15.7682 & 5.3   & 2.96  & 4.13\pm0.28 & -0.658\pm0.034 & 0.2986\pm0.0581 & -0.625\pm0.037 & -30.18 & -16.45 \\
13    & 5.9952 & 15.7684 & 4.3   & 5.90  & 4.16\pm0.22 & -0.610\pm0.020 & 0.1732\pm0.0420 & -0.725\pm0.025 & -30.31 & -16.45 \\
14    & 5.9977 & 15.7686 & 2.3   & 3.52  & 3.90\pm0.19 & -0.470\pm0.026 & 0.1633\pm0.0397 & -0.570\pm0.030 & -6.32 & -3.24 \\
15    & 5.9935 & 15.7696 & 6.1   & 12.79 & 4.39\pm0.42 & -1.104\pm0.059 & 1.3145\pm0.0749 & -0.940\pm0.068 & -66.76 & -17.18 \\
16    & 5.9922 & 15.7714 & 8.1   & 19.80 & 4.46\pm0.10 & -0.954\pm0.019 & 1.3544\pm0.0261 & -0.725\pm0.016 & -50.78 & -16.78 \\
17    & 5.9994 & 15.7716 & 3.2   & 10.49 & 4.34\pm0.14 & -0.629\pm0.024 & 0.2426\pm0.0356 & -0.661\pm0.024 & 38.13 & 13.55 \\
18    & 5.9920 & 15.7753 & 10.8  & 1.60  & 3.60\pm0.24 & -1.018\pm0.042 & 1.4242\pm0.0565 & -0.562\pm0.088 & 15.93 & -6.40 \\
19    & 5.9946 & 15.7744 & 7.2   & 24.62 & 4.23\pm0.19 & -1.107\pm0.027 & 1.5927\pm0.0338 & -0.956\pm0.024 & -35.09 & 3.64 \\
20    & 6.0028 & 15.7742 & 9.3   & 2.12  & 4.44\pm0.41 & -0.560\pm0.026 & 0.6094\pm0.0555 & -0.414\pm0.019 & 41.59 & 32.41 \\
21    & 5.9926 & 15.7763 & 11.1  & 1.64  & 3.88\pm0.30 & -0.898\pm0.054 & 0.9573\pm0.0772 & -0.533\pm0.046 & -1.43 & 2.89 \\
22    & 5.9980 & 15.7763 & 8.3   & 1.56  & 3.90\pm0.30 & -0.727\pm0.038 & 0.9263\pm0.0517 & -0.546\pm0.037 & 73.33 & 21.22 \\
23    & 6.0008 & 15.7776 & 11.0  & 9.27  & 4.25\pm0.14 & -0.998\pm0.016 & 1.4131\pm0.0227 & -0.801\pm0.018 & 32.00 & 30.57 \\
24    & 6.0007 & 15.7816 & 16.0  & 2.37  & 4.12\pm0.19 & -1.212\pm0.052 & 1.7082\pm0.0563 & -0.842\pm0.035 & 34.68 & 37.89 \\
25    & 6.0020 & 15.7819 & 17.1  & 2.40  & 3.85\pm0.16 & -1.070\pm0.033 & 1.2779\pm0.0429 & -0.638\pm0.023 & 55.76 & 42.14 \\
26    & 6.0034 & 15.7825 & 18.6  & 1.09  & 3.97\pm0.20 & -1.122\pm0.085 & 1.6681\pm0.0901 & -0.638\pm0.038 & 48.44 & 47.21
\enddata
\tablenotetext{}{Flux is in units of $\times 10^{-16}~ \rm erg ~s^{-1} ~cm^{-2}$.}
\end{deluxetable*}

\begin{deluxetable*}{CCCCCC}
\tablenum{4}
\label{tab:linearfits}
\tablecaption{Metallicity Radial Gradients}
\tablewidth{0pt}
\tabletypesize{\footnotesize}
\tablehead{
\colhead{y} & \colhead{x} &  \colhead{slope m} & \colhead{intercept b} & \colhead{comment} & \colhead{Spearman $\rho$}
}
\startdata
\rm{N2} & \rm R & -0.032\pm0.007 & -0.564\pm0.073 & &\\
\rm{12 + log(O/H)_{N2}} & \rm R & -0.017\pm0.004 & 8.646\pm0.038 &  & -0.60 \\
\rm{12 + log(O/H)_{N2}} & \rm R & -0.020\pm0.003 & 8.693\pm0.029 & \text{ALMs removed} & -0.80 \\
\rm{12 + log(O/H)_{N2}} & \rm R/R_{25} & -0.191\pm0.041 & 8.644\pm0.038 & &\\
\rm{O3N2} & \rm  R & 0.083\pm0.015 & 0.249\pm0.166 & &\\
\rm{12 + log(O/H)_{O3N2}} & \rm  R & -0.020\pm0.004 & 8.703\pm0.041 & &-0.63\\
\rm{12 + log(O/H)_{O3N2}} & \rm  R & -0.023\pm0.003 & 8.746\pm0.035 & \text{ALMs removed} &  -0.77
\enddata

\tablenotetext{}{Linear fit equations to the metallicity versus galactocentric radius data. Slopes are all in dex kpc$^{-1}$ except for when x = $\rm R/R_{25}$ which gives slope units of dex $\rm R_{25}^{-1}$. Intercepts are all in dex.}
\end{deluxetable*}

\begin{deluxetable*}{CCCCCCCCCCCC}
\rotate
\tablenum{5}
\label{tab:FMR}
\tablecaption{Calculated Physical Properties of \HII Regions}
\tablewidth{0pt}
\tabletypesize{\footnotesize}
\tablehead{
\colhead{\HII} & \colhead{Radius} & \colhead{$\rm 12+log(O/H)_{N2}$} &  \colhead{$\rm 12+log(O/H)_{O3N2}$} & \colhead{log($\rm \Sigma_{SFR}$)} & \colhead{$\rm log(\Sigma_\star)$} & \colhead{$\rm log(M_{HI})$} &  \colhead{$f_{gas}$} & \colhead{$\rm log(y_{eff})$} & \colhead{$\eta$} & \colhead{$\eta$} & \colhead{$\eta$}\\
\colhead{Region} & \colhead{(kpc)} & \colhead{} & \colhead{} & \colhead{($\rm M_\odot~yr^{-1}~kpc^{-2}$)} & \colhead{($\rm M_\odot~kpc^{-2}$)} & \colhead{($\rm M_\odot$)} & \colhead{} & \colhead{} & \colhead{for $\rm y_O=0.009$} & \colhead{for $\rm y_O=0.014$} & \colhead{for $\rm y_O=0.037$}
}
\startdata
1     & 14.4  & 8.43\pm0.07 & 8.456\pm0.036 & -2.79 & 7.02  & 6.38  & 0.47  & -2.38 & 1.72  & 3.40  & 10.70 \\
2     & 14.4  & 8.43\pm0.05 & 8.405\pm0.030 & -2.49 & 7.09  & 6.43  & 0.46  & -2.39 & 1.73  & 3.40  & 10.70 \\
3     & 11.7  & 8.52\pm0.03 & 8.565\pm0.014 & -2.64 & 7.08  & 6.43  & 0.46  & -2.30 & 1.83  & 2.54  & 8.51 \\
4     & 12.2  & 8.43\pm0.04 & 8.392\pm0.024 & -2.26 & 7.29  & 6.42  & 0.34  & -2.53 & 1.98  & 3.43  & 10.70 \\
5     & 13.6  & 8.46\pm0.01 & 8.430\pm0.010 & -2.21 & 7.21  & 6.37  & 0.36  & -2.48 & 1.26  & 3.13  & 9.92 \\
6     & 11.4  & 8.46\pm0.01 & 8.465\pm0.010 & -1.77 & 7.68  & 6.45  & 0.19  & -2.69 & 2.34  & 3.13  & 9.92 \\
7     & 11.0  & 8.41\pm0.01 & 8.394\pm0.008 & -0.99 & 8.12  & 6.45  & 0.08  & -2.93 & 1.78  & 3.64  & 11.25 \\
8     & 8.3   & 8.51\pm0.02 & 8.533\pm0.012 & -2.43 & 7.56  & 6.51  & 0.25  & -2.56 & 0.88  & 2.68  & 8.73 \\
9     & 6.1   & 8.53\pm0.02 & 8.567\pm0.013 & -2.44 & 7.75  & 6.53  & 0.19  & -2.63 & 2.34  & 2.52  & 8.29 \\
10    & 10.0  & 8.51\pm0.02 & 8.525\pm0.007 & -2.05 & 7.56  & 6.48  & 0.24  & -2.57 & 1.39  & 2.68  & 8.73 \\
11    & 9.1   & 8.53\pm0.01 & 8.544\pm0.011 & -1.71 & 7.87  & 6.49  & 0.14  & -2.70 & 1.91  & 2.52  & 8.29 \\
12    & 5.3   & 8.59\pm0.02 & 8.686\pm0.012 & -2.38 & 7.92  & 6.53  & 0.14  & -2.64 & 2.84  & 2.06  & 7.09 \\
13    & 4.3   & 8.62\pm0.01 & 8.712\pm0.009 & -2.07 & 8.12  & 6.53  & 0.09  & -2.69 & 2.42  & 1.86  & 6.55 \\
14    & 2.3   & 8.70\pm0.02 & 8.714\pm0.009 & -2.26 & 8.31  & 6.54  & 0.20  & -2.44 & 2.18  & 1.38  & 5.28 \\
15    & 6.1   & 8.36\pm0.03 & 8.444\pm0.021 & -1.80 & 7.96  & 6.51  & 0.12  & -2.90 & 1.76  & 4.20  & 12.75 \\
16    & 8.1   & 8.44\pm0.01 & 8.433\pm0.008 & -1.59 & 7.89  & 6.48  & 0.13  & -2.80 & 0.97  & 3.33  & 10.43 \\
17    & 3.2   & 8.61\pm0.02 & 8.698\pm0.007 & -1.82 & 8.26  & 6.52  & 0.22  & -2.51 & 1.36  & 1.92  & 6.73 \\
18    & 10.8  & 8.41\pm0.02 & 8.414\pm0.017 & -2.69 & 6.99  & 6.43  & 0.51  & -2.34 & 1.26  & 3.58  & 11.25 \\
19    & 7.2   & 8.36\pm0.02 & 8.364\pm0.010 & -1.51 & 7.94  & 6.49  & 0.12  & -2.90 & 1.12  & 4.20  & 12.75 \\
20    & 9.3   & 8.65\pm0.02 & 8.619\pm0.013 & -2.51 & 7.66  & 6.46  & 0.19  & -2.50 & 1.66  & 1.67  & 6.05 \\
21    & 11.1  & 8.47\pm0.03 & 8.537\pm0.019 & -2.67 & 7.04  & 6.43  & 0.48  & -2.32 & 0.84  & 2.98  & 9.67 \\
22    & 8.3   & 8.56\pm0.02 & 8.544\pm0.013 & -2.67 & 7.65  & 6.47  & 0.20  & -2.57 & 1.37  & 2.28  & 7.67 \\
23    & 11.0  & 8.42\pm0.01 & 8.417\pm0.007 & -1.93 & 7.59  & 6.44  & 0.21  & -2.70 & 1.11  & 3.53  & 10.97 \\
24    & 16.0  & 8.30\pm0.03 & 8.328\pm0.018 & -2.54 & 7.24  & 6.35  & 0.33  & -2.68 & 0.71  & 4.97  & 14.78 \\
25    & 17.1  & 8.38\pm0.02 & 8.454\pm0.012 & -2.52 & 7.18  & 6.34  & 0.36  & -2.56 & 0.52  & 3.97  & 12.13 \\
26    & 18.6  & 8.35\pm0.05 & 8.341\pm0.029 & -2.87 & 7.44  & 6.28  & 0.21  & -2.77 & 1.63  & 4.32  & 13.07 
\enddata
\end{deluxetable*}

\begin{deluxetable}{lC}
\tablenum{6}
\label{tab:companion properties}
\tablecaption{Properties of SDSS J002353.17+154356.8}
\tablewidth{0pt}
\tabletypesize{\small}
\tablehead{
\colhead{Parameter} & \colhead{Value}
}
\startdata
{\rm R.A.}\tablenotemark{a}  ($\alpha_{2000}$) & 00^h23^m53^{s}.17\\
{\rm Decl.}\tablenotemark{a} ($\delta_{2000}$) & +15\degree43^{\prime}56^{\prime\prime}.87\\
{\rm Redshift} & 0.01747 \\
{\rm Projected distance to NGC 99}& 57~$\rm kpc$\\
{\rm Velocity offset to NGC 99 ($\Delta V_{H\alpha}$)} & -74.51~$\rm km/s$\\
{\rm Stellar~mass}\tablenotemark{b} & $8.62\times10^{8}~M_\odot$ \\
{\rm H~I~mass}\tablenotemark{c} & $5.8\times10^{9}~M_\odot$ \\
{\rm N2} & $-$0.46 \\
{\rm 12+log(O/H)} & 8.71
\enddata
\tablenotetext{a}{\cite{2MASS_2006AJ....131.1163S}}
\tablenotetext{b}{\cite{Padave_2024ApJ...960...24P}}
\tablenotetext{c}{Gim et al., in prep.}
\end{deluxetable}




\end{document}